\documentclass[journal]{IEEEtran}
\usepackage{amsmath,amsfonts}
\usepackage{algorithmic}
\usepackage{algorithm}
\usepackage{array}
\usepackage[caption=false,font=normalsize,labelfont=sf,textfont=sf]{subfig}
\usepackage{textcomp}
\usepackage{stfloats}
\usepackage{url}
\usepackage{verbatim}
\usepackage{graphicx}
\usepackage{cite}
\begin{document}

\title{Performance Analysis of Photon-Limited Free-Space Optical Communications with Practical Photon-Counting Receivers}

\author{Chen Wang, Zhiyong Xu, Jingyuan Wang, Jianhua Li, Weifeng Mou, Huatao Zhu, Jiyong Zhao, Yang Su, Yimin Wang, and Ailin Qi
\thanks{This work was supported by National Natural Science Foundation of China (Grant No. 62271502, No. 62171463, and No. 61975238) and Natural Science Foundation of Jiangsu Province (Grant No. BK20231486). (\textit{Corresponding author: Jingyuan Wang}.) 

Chen Wang, Zhiyong Xu, Jingyuan Wang,  Jianhua Li, Jiyong Zhao, Yang Su, Yimin Wang, and Ailin Qi are with the College of Communications Engineering, Army Engineering University of PLA, Nanjing 210007, China (e-mail: 0910210239@njust.edu.cn; njxzy123@163.com; 13813975111@163.com; 18021528752@163.com; zhaojiyong@whu.edu.cn; qieziyangyang@163.com; vivhappyrom@163.com; qial1212@163.com).

Weifeng Mou, Huatao Zhu are with the College of Information and Communication, National University of Defense Technology, Wuhan 430010, China (e-mail: weifengmou@126.com; zhuhuatao2008@163.com). 

}
\thanks{Manuscript received  ,  ; revised  ,  .}}

\markboth{Journal of \LaTeX\ Class Files,~Vol.~ , No.~ ,  ~ }%
{Shell \MakeLowercase{\textit{et al.}}: A Sample Article Using IEEEtran.cls for IEEE Journals}

\IEEEpubid{}

\maketitle

\begin{abstract}
The non-perfect factors of practical photon-counting receiver are recognized as a significant challenge for long-distance photon-limited free-space optical (FSO) communication systems. This paper presents a comprehensive analytical framework for modeling the statistical properties of time-gated single-photon avalanche diode (TG-SPAD) based photon-counting receivers in presence of dead time, non-photon-number-resolving and afterpulsing effect. Drawing upon the non-Markovian characteristic of afterpulsing effect, we formulate a closed-form approximation for the probability mass function (PMF) of photon counts, when high-order pulse amplitude modulation (PAM) is used. Unlike the photon counts from a perfect photon-counting receiver, which adhere to a Poisson arrival process, the photon counts from a practical TG-SPAD based receiver are instead approximated by a binomial distribution. Additionally, by employing the maximum likelihood (ML) criterion, we derive a refined closed-form formula for determining the threshold in high-order PAM, thereby facilitating the development of an analytical model for the symbol error rate (SER). Utilizing this analytical SER model, the system performance is investigated. The numerical results underscore the crucial need to suppress background radiation below the tolerated threshold and to maintain a sufficient number of gates in order to achieve a target SER.
\end{abstract}


\section{Introduction}
\IEEEPARstart{T}{he} laser systems show promise for use in various wireless communication scenarios, such as free space optical (FSO) communication, deep space optical communication (DSOC) and underwater optical wireless communication (UOWC) \cite{ref1,ref3,ref4}. They offer a larger bandwidth, lower emitting power, and smaller antenna aperture when compared to radio frequency (RF) systems. Typically, an optical receiver detects the received signals using a conventional continuous waveform receiver, such as a p-i-n diode and avalanche photodiode (APD). These detectors have simple structures and stable performances. However, in photon-limited optical communications, the received signals often fall below the sensitivity of p-i-n diodes and APDs, and lost in the intrinsic thermal noise. To enhance sensitivity in such challenging conditions, single-photon detector and photon-counting technology are being utilized \cite{ref9,ref10}. The single-photon avalanche diode (SPAD) is particularly noteworthy for its extremely high avalanche gain, which allows it to overcome gain-dependent excess noise and intrinsic thermal noise \cite{ref11}. Consequently, the SPAD is capable of detecting single-photon signals. Due to its single-photon sensitivity, the SPAD is receiving growing interest for use in FSO communication, especially in channels experiencing deep fading \cite{ref12,ref13}.

Based on different quenching mechanisms, there are three principal operation modes: passive quenching (PQ), active quenching (AQ), and time-gated (TG). PQ and AQ circuits initiate the quenching of SPADs when the avalanche current surpasses a specified threshold by reducing the bias voltage. However, this leads to a huge avalanche current flowing through the p-n junction and exacerbates the afterpulsing effect \cite{ref14}. In addition, when dead time is less than symbol duration, PQ- and AQ-SPAD based photon-counting receivers suffer from inter-symbol interference (ISI) due to the Poisson point process of arrival instances of photons \cite{ref22}. In contrast, TG-SPAD, which employ an external periodic ultrashort gating signal to lower the bias voltage, can shorten the quenching time to the sub-nanosecond (ns) order. This controls the avalanche current at a relatively lower level. Therefore, TG-SPADs are preferred for achieving high detection bandwidth with a reduced afterpulsing effect \cite{ref23}. Furthermore, ISI induced by dead time is eliminated owing to the periodic gating signals \cite{ref24,ref25}. By using sinusoidal gating and self-differencing techniques, TG-SPADs have achieved detection bandwidth exceeding GHz \cite{ref26,ref27}. Consequently, TG-SPADs have great potential to realize high-bandwidth photon-counting receivers with a low afterpulsing effect.

\subsection{Related Works}
For photon-counting receivers, the received signals are typically characterized by discrete photoelectrons, which follow a Poisson distribution. In such a Poisson channel, the traditional additive white Gaussian noise (AWGN) model is no longer applicable. Early studies often presumed a perfect photon-counting receiver and focused on the Poisson distribution of the photon counts \cite{ref28,ref29}. However, achieving a perfect photon-counting receiver is challenging, and the photon counts during symbol duration are non-linearly distorted due to the dead time and non-photon-number-resolving \cite{ref30,ref31}. 

Recent studies \cite{ref32,ref33,ref34} have introduced practical TG-SPAD based photon-counting receivers for long-distance FSO systems and accounted for the blocking losses caused by dead time. Nonetheless, the presupposed Poisson statistics for photon counts fall short of accuracy. In our previous work \cite{ref35}, we explored the statistical model for TG-SPAD considering both dead time and non-photon-number-resolving. Furthermore, the researches \cite{ref36,ref37,ref38} confirmed the use of non-photon-number-resolving TG-SPAD receivers with pulse position modulation (PPM) and introduced a binomial distribution to more accurately depict photon count statistics, offering better predictions for system error performance than the Poisson model. Studies on FSO systems that utilize TG-SPAD arrays to reduce ISI caused by dead time were investigated in \cite{ref25,ref39,ref40}, and the photon count statistics were approximated using a Gaussian distribution by central limit theorem. In addition to the Poisson process of photon arrivals, the afterpulsing effect is dependent on the history of events, and introduces further variability in detected photon counts \cite{ref41,ref42}. However, the above studies did not take into account afterpulsing effect.

In our previous work \cite{ref43}, we investigated the statistical characteristics of an InGaAs/InP TG-SPAD receiver and  the impact of afterpulsing effect on FSO systems, and derived an approximate probability distribution for the photon counts with on-off keying (OOK) modulation. Moreover, we extended this analysis to demonstrate that communication performance can be enhanced by balancing the afterpulsing effect, dead time, and optical attenuation control \cite{ref44,ref45}. This model was also applied to study the InGaAs/InAlAs SPAD receiver, which also suffers from afterpulsing effect \cite{ref46}. Nevertheless, for high-order modulation scenarios, such as pulse amplitude modulation (PAM) and orthogonal frequency division multiplexing (OFDM), the precision of this model diminishes sharply. Consequently, there is an urgent need for statistical models that account for the afterpulsing effect in practical photon-counting receivers with high-order modulation to accurately assess error performance and optimize receiver parameters. 

\subsection{Our Contributions}
The main contributions can be summarized as follows:
\subsubsection{Afterpulsing Effect Formulation} 
We investigate the non-Markovian afterpulsing effect from the perspective of stochastic process theory. Unlike previous work \cite{ref47}, which analyzed the finite release time of trapped carriers in terms of a Markovian process, we reveal the fundamental trigger probability linked to the buildup of afterpulses, and discuss how different parameters can enhance or impair communication performance, including signal and background photon rates, the number of gates and array scale.

\subsubsection{Statistical Modeling of Photon Counts}
For high-order PAM signals with a constant photon rate during the symbol duration, we propose an approximation method to estimate the trigger probability for single SPAD and SPAD array. We validate the accuracy of the proposed model for the probability distribution of photon counts by comparison with a Monte Carlo simulation that includes non-Markovian afterpulsing effect. Additionally, we investigate and contrast the SER in both single SPAD and SPAD array configurations.

\subsubsection{Photon-Counting Signal Estimation and Decision Scheme}
Based on the photon counts distribution, we propose a general signal estimation scheme along with a specialized history-enhanced estimation scheme. We formulate the signal estimation equations and offer explicit threshold calculation formulas. Our study also delineates the relationship between the SER and various parameters of the photon-counting receiver. Furthermore, we demonstrate that a more pronounced afterpulsing effect invariably compromises communication performance in high-order PAM systems. This finding contrasts with earlier research \cite{ref43}, which characterized afterpulse as improving error performance in low optical intensity conditions for OOK systems.

To the best of our knowledge, no existing analytical work has provided closed-form expressions for the probability distribution of photon counts, which take into account dead time, non-photon-number-resolving, and afterpulsing effect. The approximation method presented in this paper is versatile and can be applied to various types of single-photon detectors operating in a similar mode.

\subsection{Organization}
The remainder of this paper is organized as follows. Section \uppercase\expandafter{\romannumeral2} explains the operational principle of TG-SPADs and the detection scheme of practical photon-counting receivers. Section \uppercase\expandafter{\romannumeral3} provides the probability distribution of photon counts based on the concepts of non-Markovian afterpulsing effect and verifies its accuracy by comparing it with a Monte Carlo simulation. In Section \uppercase\expandafter{\romannumeral4}, we propose a general signal estimation scheme as well as a specialized history-enhanced estimation scheme. Section \uppercase\expandafter{\romannumeral5} presents numerical results and discussions on communication performance. Finally, Section \uppercase\expandafter{\romannumeral6} contains the concluding remarks.

\section{Photon-Counting Receiver}
\subsection{TG-SPAD Operation}

\begin{figure}[!t]
\centering
\includegraphics[width=2.5in]{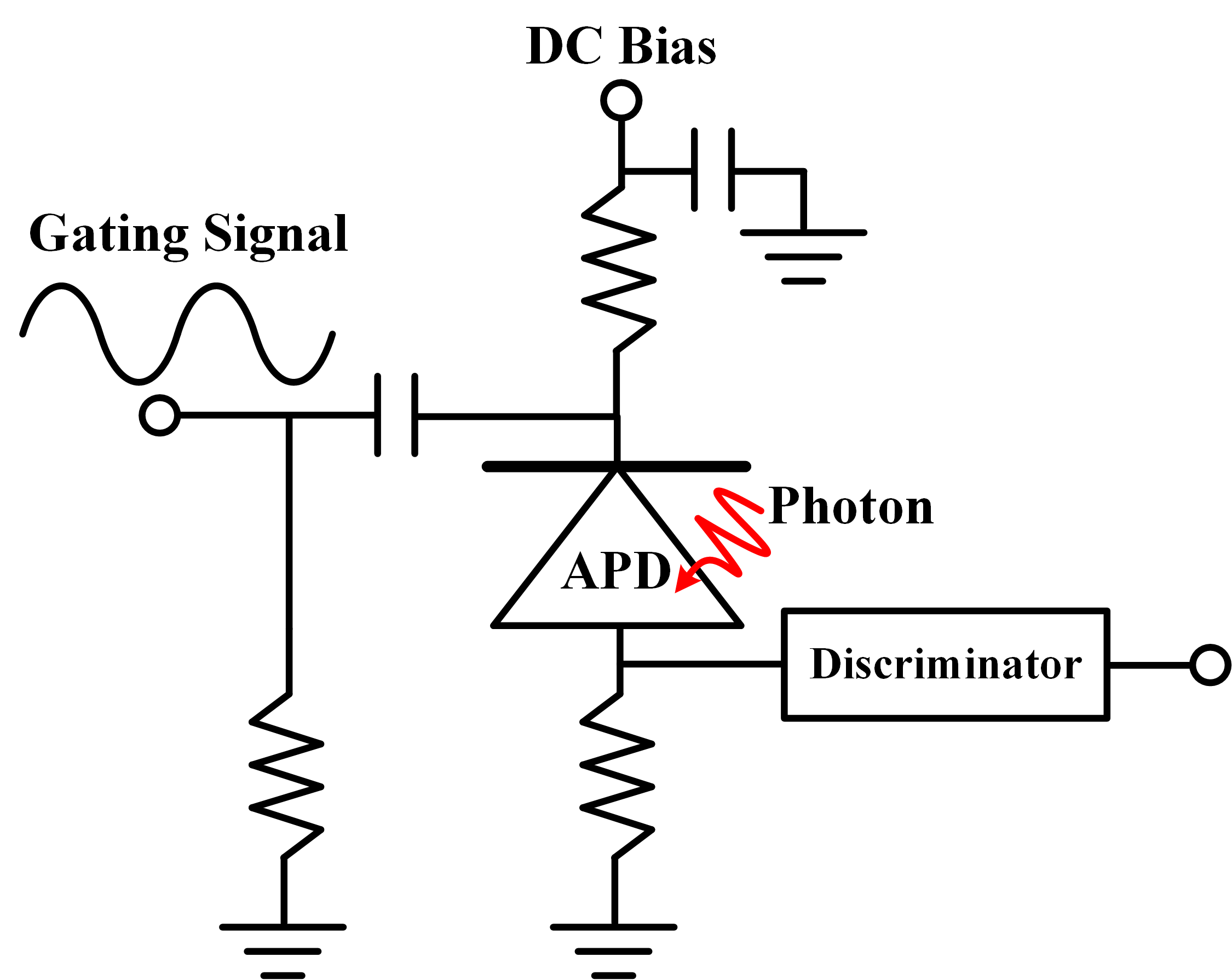}
\caption{TG-SPAD circuit using a sinusoidal gating signal.}
\label{fig_1}
\end{figure}

As depicted in Fig. 1, we present the time-gated circuit that employs a sinusoidal gating signal. The APD module serves as a single-photon detection element and is supplied with a reverse bias voltage. After amplification by a high-power amplifier, the signal generator produces a sinusoidal voltage that functions as the gating signal. The avalanche events from the APD module are then sent to the discriminator. Within the discriminator, an ultra-high-speed comparator circuit is used to differentiate the signals \cite{ref27}. This process results in a logic output at the emitter-coupled-logic level.

The specific operational principle of the TG-SPAD is illustrated in the red dashed box in Fig. 2. During time-gated mode operation, the SPAD is regularly armed and can only record whether an avalanche event occurs during a given window, referred to as the gate-ON interval ${\tau _{\rm{g}}}$. The amplitude of avalanche current is strongly influenced by gate-ON interval. To suppress afterpulsing effect, the gate-ON interval is typically shortened to the order of nanoseconds. Within a gate, the detector is able to complete a single detection and produce an output for the avalanche event. Conversely, the SPAD is inactive during gate-OFF interval ${\tau _{\rm{d}}}$. Additionally, the detected photon counts represent the number of avalanche events within the symbol duration ${T_{\rm{s}}}$. The detection cycle ${\tau _{{\rm{cyc}}}} \buildrel \Delta \over = {\tau _{\rm{g}}} + {\tau _{\rm{d}}}$, is defined as the sum of the gate-ON interval and the gate-OFF interval.

\subsection{Detection Scheme}

\begin{figure}[!t]
\centering
\includegraphics[width=3.4in]{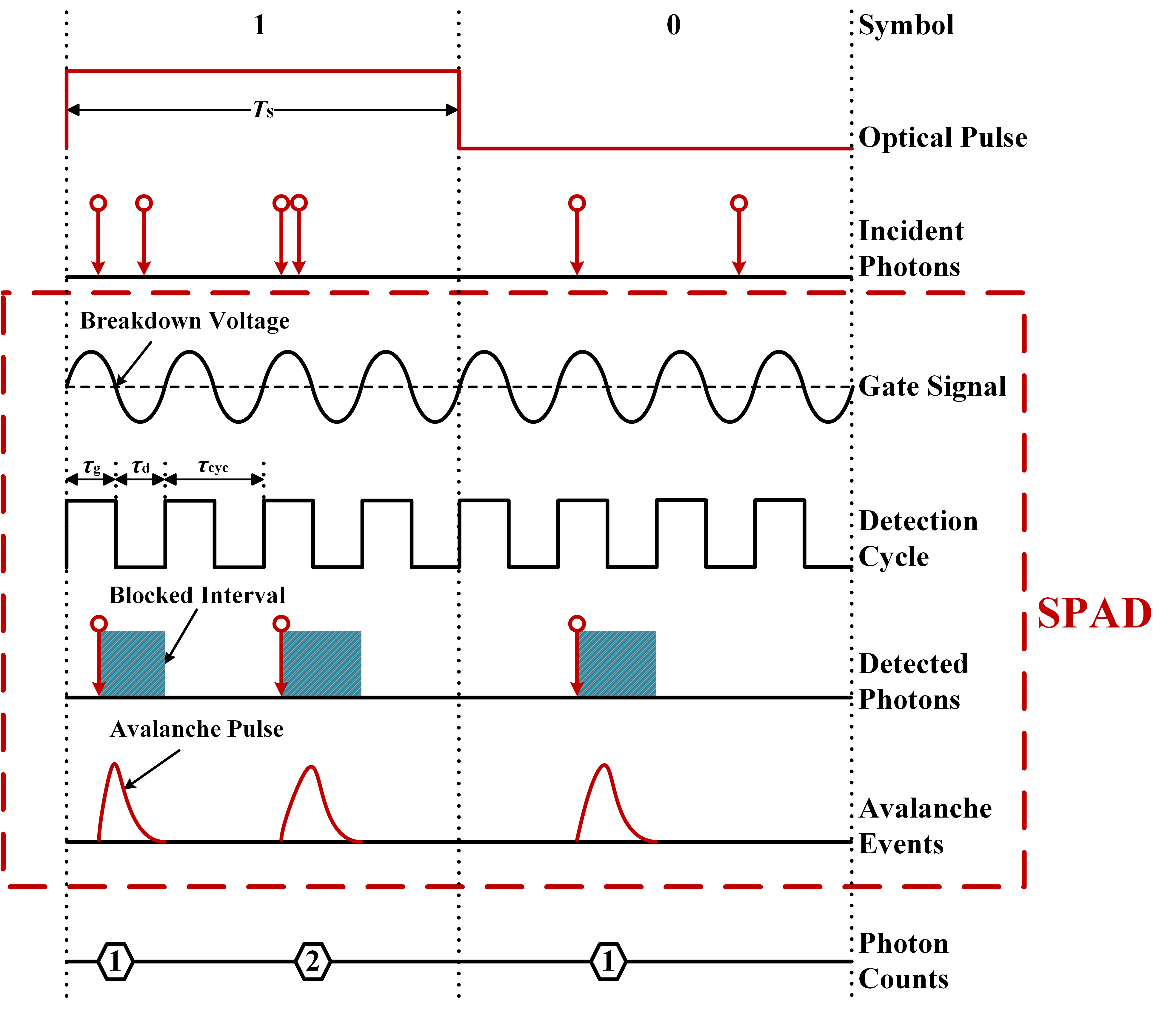}
\caption{The detection scheme for a single SPAD based photon-counting receiver.}
\label{fig_2}
\end{figure}

\begin{figure}[!t]
\centering
\includegraphics[width=3.4in]{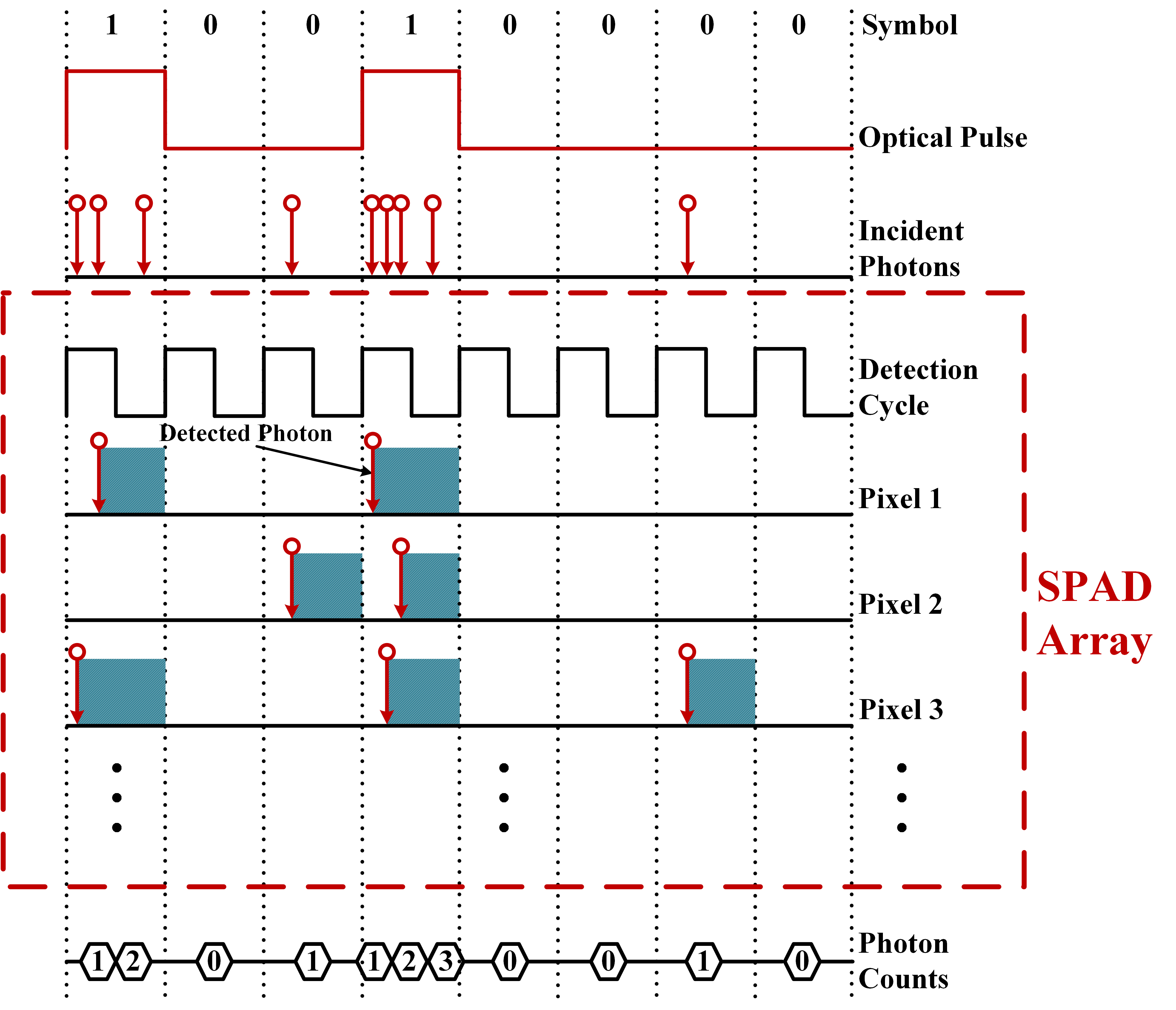}
\caption{The detection scheme for a SPAD array based photon-counting receiver.}
\label{fig_3}
\end{figure}

In addition to optical signals, background radiation and afterpulses can also contribute to false photon counts. To overcome the randomness of a single gate output, symbol information is determined by opening several gates within the symbol duration and comparing the detected photon counts with a threshold. As discussed in our previous research \cite{ref35}, achieving a lower SER is possible by opening more gates within the symbol duration. However, opening more gates will decrease the information transfer rate.

For a single SPAD based photon-counting receiver, the detection scheme is depicted in Fig. 2. The TG-SPAD opens a series of gates during the symbol duration, and the resulting photon counts are recorded. Due to the dead time, only a fraction of the signal photons are detected. Consequently, the continuous pulses are converted into a sequence of discrete avalanche events after detection. The signal intensity is then estimated based on the number of avalanche events (photon counts) within the symbol duration. Due to the non-photon-number-resolving, the photon counts are non-linearly distorted at medium and high optical intensity levels. To maintain an appropriate dynamic range and communication performance, the number of gates is typically on the order of several hundred.

To open more gates without compromising the information transfer rate, SPAD pixels are fabricated into an array for optical signal detection. Compared to individual SPADs, SPAD arrays can handle more detection events within the symbol duration. Utilizing SPAD arrays can enhance communication performance without the drawback of a decreased information transfer rate. As illustrated in Fig. 3, the SPAD array based photon-counting receiver outputs the photon counts in a detection cycle, akin to an analog-to-digital converter with single-photon sensitivity. Due to the different detection schemes, the afterpulsing effect on single SPADs and SPAD arrays is not identical, resulting in different formulas for calculating trigger probability. The trigger probability is the likelihood of incident photons initiating an avalanche event and producing a photon count during the gate-ON interval.

\section{Photon-Counting Statistics}
\subsection{Non-Markovian Afterpulsing Effect}

\begin{figure}[!t]
\centering
\includegraphics[width=2.5in]{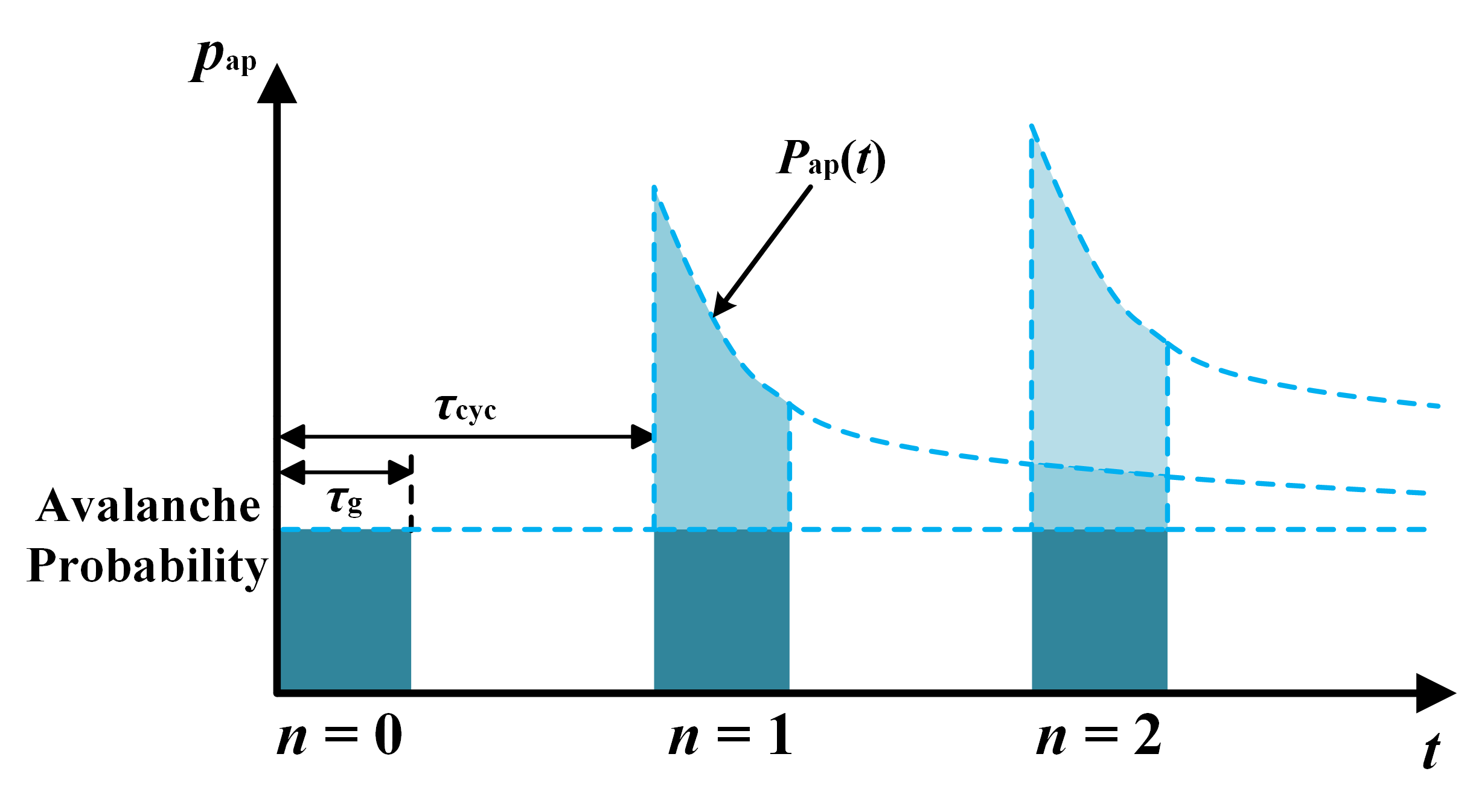}
\caption{The diagram illustrating the non-Markovian afterpulsing effect.}
\label{fig_4}
\end{figure}

During an afterpulsing event, trapped charge carriers are released following a random time interval determined by the trap lifetime. The trap lifetime is influenced by the energy level and the properties of material \cite{ref51}. To describe the afterpulsing probability (AP), a multi-exponential model, as introduced in \cite{ref52}, is employed. AP is defined as the likelihood of carriers, captured during a pre-avalanche event, being released in the subsequent gate, thereby triggering a spurious avalanche event. The probability density function (PDF) for an afterpulsing event at a specific time instant $t$ is given by \cite{ref52}
\begin{equation}\label{eq1}
f(t) = \sum\nolimits_j {{A_j}{{\rm{e}}^{ - {t \mathord{\left/
 {\vphantom {t {{\tau _{{\rm{rel,}}}}_j}}} \right.
 \kern-\nulldelimiterspace} {{\tau _{{\rm{rel,}}}}_j}}}}} 
\end{equation}
where ${A_j}$ denotes the amplitude of exponential component related to trap density, and ${\tau _{{\rm{rel, }}j}}$ represents the lifetime of the $j$-th trap. A trap is a defect within the material that can capture a carrier during an avalanche event.

As depicted in Fig. 4, the incidence of consecutive avalanche events can markedly increase the number of trapped carriers, thereby enhancing the AP. Consequently, the AP is not solely determined by the characteristics of the traps but is also connected to the sequence of previous avalanche events. The AP is a cumulative process involving the release of carriers from all preceding avalanche events, which demonstrates non-Markovian behavior \cite{ref53}. We denote the $n$-order AP, corresponding to the  $n$-th gate following a primary avalanche event, as ${p_{{\rm{ap}}}}\left( n \right) \buildrel \Delta \over = \int_{n{\tau _{{\rm{cyc}}}}}^{n{\tau _{{\rm{cyc}}}} + {\tau _{\rm{g}}}} {f\left( t \right){\rm{d}}t}$. According to (\ref{eq1}), the $n$-order AP is given by
\begin{equation}\label{eq2}
{p_{{\rm{ap}}}}\left( n \right) = \int_{n{\tau _{{\rm{cyc}}}}}^{n{\tau _{{\rm{cyc}}}} + {\tau _{\rm{g}}}} {\left( {\sum\nolimits_j {{A_j}{{\rm{e}}^{ - {t \mathord{\left/
 {\vphantom {t {{\tau _{{\rm{rel,}}}}_j}}} \right.
 \kern-\nulldelimiterspace} {{\tau _{{\rm{rel,}}}}_j}}}}} } \right){\rm{d}}t}
\end{equation}

If the trap lifetime significantly exceeds the detection cycle, the afterpulsing effect manifests as a stable cumulative process and can be regarded as part of the intrinsic dark carriers. Conversely, if the trap lifetime is much shorter than the detection cycle, most trapped carriers are released during the dead time and can be ignored. However, when the trap lifetime is on the order of the detection cycle, trapped carriers may persist beyond the dead time and become free in the subsequent gates. For fabricated SPADs, the energy levels of the traps are determined during diode fabrication, and these characteristics have been thoroughly investigated. Hence, we do not consider the time-dependent distribution of AP. Accurate values for trap lifetimes and the amplitudes of exponential components are ascertainable via experimental procedures.  

In the following section, we will explore the probability mass function (PMF) of photon counts, when photon-counting receivers continuously detect high-order PAM signals with pseudo-random transmitting symbols.
\subsection{PMF of Photon Counts}
Define the avalanche probability as the likelihood of initiating an avalanche event by incident photons without the influence of afterpulsing effect. Since weak light photons adhere to a Poisson arrival process ${\mathbb{P}}\left( \lambda  \right)$, the avalanche probabilities across different gates are independent identically distributed. Let ${p_n}$ represent the avalanche probability for the $n$-th gate, which is expressed as follows \cite{ref35}:
\begin{equation}\label{eq3}
{p_n} = 1 - {\operatorname{e} ^{ - {\lambda _n}}}
\end{equation}
where ${\lambda _n}$ signifies the average number of carriers in the $n$-th gate, defined as $
{\lambda _n} \buildrel \Delta \over = \left( {{\lambda _{\rm{s}}}_{,n} + {\lambda _{\rm{b}}}} \right){p_{{\rm{de}}}}{\tau _{\rm{g}}} + {\lambda _{\rm{d}}}{\tau _{\rm{g}}}$. Here, ${\lambda _{\text{s}}}_{,n},{\lambda _{\text{b}}},{\lambda _{\text{d}}}$  denote the signal photon rate, background photon rate and dark carrier rate in the $n$-th gate, respectively, and ${p_{{\text{de}}}}$ represents the photon detection efficiency (PDE) of the SPAD.

Define the trigger probability as the likelihood of an avalanche event being triggered by incident photons, background radiation, dark carriers, and afterpulsing effect. The complete formula for the trigger probability of the $n$-th gate ${P_n}$ is quite complex, and is contingent upon the states of all prior gates. For the initial gate, the afterpulsing effect is absent. If the $n$-th gate is activated, the trapped carriers will influence all subsequent gates. The comprehensive formulations for the trigger probabilities of the first, second, and third gates are derived as follows:

\begin{equation}\label{eq4}
\begin{array}{c}
{P_1} = {p_1} \\
{P_2} = {p_1}\left[ {{p_2} + \left( {1 - {p_2}} \right){p_{{\text{ap}}}}(1)} \right] + \left( {1 - {p_1}} \right){p_2} \\
{P_3} = {p_1}\left[ {{p_2} + \left( {1 - {p_2}} \right){p_{{\rm{ap}}}}(1)} \right]\left[ {{p_3}\left( {1 - {p_{{\rm{ap}}}}(1)} \right)} \right. \\
 \times \left. {\left( {1 - {p_{{\rm{ap}}}}(2)} \right) + 1 - \left( {1 - {p_{{\rm{ap}}}}(1)} \right)\left( {1 - {p_{{\rm{ap}}}}(2)} \right)} \right] \\
 + {p_1}\left[ {{p_2} + \left( {1 - {p_2}} \right){p_{{\rm{ap}}}}(1)} \right]\left[ {{p_3} + \left( {1 - {p_3}} \right){p_{{\rm{ap}}}}(2)} \right] \\
 + \left( {1 - {p_1}} \right){p_2}\left[ {{p_3} + \left( {1 - {p_3}} \right){p_{{\rm{ap}}}}(1)} \right] \\
 + \left( {1 - {p_1}} \right)\left( {1 - {p_2}} \right){p_3}
\end{array}
\end{equation}

Employing the approximate closed-form expression for the trigger probability that has been deduced, ${P_n}$ is determined by the historical sequence of all preceding avalanche events. As depicted in (\ref{eq4}), the complexity of ${P_n}$ escalates rapidly. For typical SPADs, the first-order AP ${p_{{\rm{ap}}}}\left( 1 \right)$ is always less than $20\% $. According to (\ref{eq2}), it can be deduced that the high-order APs $\left\{ {{p_{{\rm{ap}}}}\left( 2 \right),{p_{{\rm{ap}}}}\left( 3 \right),...,{p_{{\rm{ap}}}}\left( n \right)} \right\}$  diminish exponentially. Consequently, only the first-order terms of ${p_{{\rm{ap}}}}\left( n \right)$ are retained, and higher-order terms can be disregarded, such as ${p_{{\rm{ap}}}}\left( i \right){p_{{\rm{ap}}}}\left( j \right) \cong 0,\left( {i = 1,2,...,n,j = 1,2,...,n} \right)$. Here, a closed-form approximate expression is derived as
\begin{equation} \label{eq5}
{P_n} = {p_n} + \left( {1 - {p_n}} \right)\sum\limits_{i = 1}^{n - 1} {{p_i}{p_{{\rm{ap}}}}(n - i)} 
\end{equation}The detailed proof of (\ref{eq5}) is provided in Appendix A.

For an arbitrary symbol duration in continuous operation receiver, the trigger probability can be expressed as $\mathop {\lim }\limits_{n \to \infty } {P_n}$. By inserting (\ref{eq2}) into (\ref{eq5}), the asymptotic expression for the trigger probability is obtained as
\begin{align} \label{eq6}
\begin{array}{c}
\mathop {\lim }\limits_{n \to \infty } {P_n} = {p_n} + \left( {1 - {p_n}} \right)\mathop {\lim }\limits_{n \to \infty } \sum\limits_{i = 1}^{n - 1} {{p_i}{p_{{\rm{ap}}}}(n - i)} \\
= {p_n} + {p_{\rm{a}}}\left( {1 - {p_n}} \right)\sum\nolimits_j {\left( {{A_j}{\tau _{{\rm{rel,}}}}_j\frac{{{{\rm{e}}^{{{{\tau _{\rm{g}}}} \mathord{\left/
 {\vphantom {{{\tau _{\rm{g}}}} {{\tau _{{\rm{rel,}}}}_j}}} \right.
 \kern-\nulldelimiterspace} {{\tau _{{\rm{rel,}}}}_j}}}} - 1}}{{{{\rm{e}}^{{{{\tau _{\rm{g}}}} \mathord{\left/
 {\vphantom {{{\tau _{\rm{g}}}} {{\tau _{{\rm{rel,}}}}_j}}} \right.
 \kern-\nulldelimiterspace} {{\tau _{{\rm{rel,}}}}_j}}}}\left( {{{\rm{e}}^{{{{\tau _{{\rm{cyc}}}}} \mathord{\left/
 {\vphantom {{{\tau _{{\rm{cyc}}}}} {{\tau _{{\rm{rel,}}}}_j}}} \right.
 \kern-\nulldelimiterspace} {{\tau _{{\rm{rel,}}}}_j}}}} - 1} \right)}}} \right)} 
\end{array}
\end{align}
where ${p_{\rm{a}}}$ is the equivalent avalanche probability for all previous gates ranging from $\min \left( {{p_1},{p_2},...,{p_n}} \right) \le {p_{\rm{a}}} \le \max \left( {{p_1},{p_2},...,{p_n}} \right)$. Refer to Appendix B for the comprehensive proof of (\ref{eq6}).

Assuming the signal photon rate, background photon rate, and dark carrier rate remain constant during the symbol duration, let $\chi  = \left\{ {{x_1},{x_2}, \ldots ,{x_m}, \ldots ,{x_M}} \right\}$ represent the transmitter constellation for an $M$-PAM. Based on (\ref{eq6}), let $P\left( {{x_m}} \right) \buildrel \Delta \over = \mathop {\lim }\limits_{n \to \infty } {P_n}$  denote the equivalent trigger probability for an arbitrary gate during the symbol duration of $''{{x}_{m}}''$, which can be represented as
\begin{equation} \label{eq7}
P\left( {{x_m}} \right) = p\left( {{x_m}} \right) + C{p_{\rm{a}}}\left( {1 - p\left( {{x_m}} \right)} \right)
\end{equation}
where 
\begin{equation} \nonumber
C = \sum\nolimits_j {\left( {{A_j}{\tau _{{\rm{rel,}}}}_j\frac{{{{\rm{e}}^{{{{\tau _{\rm{g}}}} \mathord{\left/
 {\vphantom {{{\tau _{\rm{g}}}} {{\tau _{{\rm{rel,}}}}_j}}} \right.
 \kern-\nulldelimiterspace} {{\tau _{{\rm{rel,}}}}_j}}}} - 1}}{{{{\rm{e}}^{{{{\tau _{\rm{g}}}} \mathord{\left/
 {\vphantom {{{\tau _{\rm{g}}}} {{\tau _{{\rm{rel,}}}}_j}}} \right.
 \kern-\nulldelimiterspace} {{\tau _{{\rm{rel,}}}}_j}}}}\left( {{{\rm{e}}^{{{{\tau _{{\rm{cyc}}}}} \mathord{\left/
 {\vphantom {{{\tau _{{\rm{cyc}}}}} {{\tau _{{\rm{rel,}}}}_j}}} \right.
 \kern-\nulldelimiterspace} {{\tau _{{\rm{rel,}}}}_j}}}} - 1} \right)}}} \right)} 
\end{equation}
where $p\left( {{x_m}} \right)$ represents the avalanche probability for an arbitrary gate during the symbol duration of $''{{x}_{m}}''$. Additionally, it is expressed by $p\left( {{x_m}} \right) = 1 - \exp \left[ { - \left( {{\lambda _{\rm{s}}}\left( {{x_m}} \right) + {\lambda _{\rm{b}}}} \right){p_{{\rm{de}}}}{\tau _{\rm{g}}} - {\lambda _{\rm{d}}}{\tau _{\rm{g}}}} \right]$, where ${\lambda _{\rm{s}}}\left( {{x_m}} \right)$ denote the signal photon rate during the symbol duration of $''{{x}_{m}}''$.

Based on the first-order moment estimation, the estimation of ${p_{\rm{a}}}$ can be calculated under the illumination of high-order PAM signals. For an arbitrary symbol sequence, ${p_{\rm{a}}}$ can be computed using the following equation:
\begin{equation} \label{eq8}
{{p}_{\text{a}}}\underset{n\to \infty }{\mathop{\lim }}\,\sum\limits_{i=1}^{n-1}{{{p}_{\text{ap}}}(i)}=\underset{n\to \infty }{\mathop{\lim }}\,\sum\limits_{i=1}^{n-1}{{{p}_{i}}{{p}_{\text{ap}}}(n-i)}
\end{equation}
where the avalanche probabilities $\left\{ {{p_1},{p_2},...,{p_{n - 1}}} \right\}$ are determined by the transmitting symbol sequence.

Due to the different detection schemes, the single SPAD receiver and SPAD array receiver have different forms of ${p_{\rm{a}}}$. Let ${p_{\rm{a}}} \buildrel \Delta \over = p_{\rm{a}}^{\rm{S}}$ denote the equivalent avalanche probability of a single SPAD, and $
{p_{\rm{a}}} \buildrel \Delta \over = p_{\rm{a}}^{\rm{A}}$  denote the equivalent avalanche probability of a SPAD array. These two schemes will be discussed separately.
For a single SPAD receiver, the SPAD detects the optical signal $N = {\textstyle{\frac{T_{s}}{{\tau _{{\rm{cyc}}}}}}}$ times during a symbol duration. The avalanche probabilities of adjacent gates are highly correlated. Based on the detection scheme of a single SPAD receiver, $p_{\rm{a}}^{\rm{S}}$ is given by
\begin{equation} \label{eq9}
p_{\rm{a}}^{\rm{S}} = \frac{{\frac{1}{N}\sum\limits_{h = 0}^{N - 1} {\left\{ {\mathop {\lim }\limits_{n \to \infty } \sum\limits_{i = 1}^{n + h - 1} {p\left( {{x_{m(n - i)}}} \right){p_{{\rm{ap}}}}(i)} } \right\}} }}{{\mathop {\lim }\limits_{n \to \infty } \sum\limits_{i = 1}^{n - 1} {{p_{{\rm{ap}}}}(i)} }}
\end{equation}
where ${x_{m(n - i)}}$ is the symbol information for the $
\left( {n - i} \right)$-th gate. Since the AP declines exponentially, the trigger probability is mainly dependent on the past several gates. The higher-order trigger probabilities can be considered nearly equal. By substituting $p\left( {{x_{m(n - i)}}} \right)$ with $p\left( {{x_m}} \right)$, the first-order moment estimation of $p_{\rm{a}}^{\rm{S}}$ is simplified to
\begin{equation} \label{eq10}
\begin{split}
p_{\rm{a}}^{\rm{S}} &\cong \frac{{\frac{1}{N}\sum\limits_{h = 0}^{N - 1} {\left\{ {\mathop {\lim }\limits_{n \to \infty } \sum\limits_{i = 1}^{n + h - 1} {p\left( {{x_m}} \right){p_{{\rm{ap}}}}(i)} } \right\}} }}{{\mathop {\lim }\limits_{n \to \infty } \sum\limits_{i = 1}^{n - 1} {{p_{{\rm{ap}}}}(i)} }}\\
 &= \frac{{p\left( {{x_m}} \right)\frac{1}{N}\sum\limits_{h = 0}^{N - 1} {\left\{ {\mathop {\lim }\limits_{n \to \infty } \sum\limits_{i = 1}^{n - 1} {{p_{{\rm{ap}}}}(i)} } \right\}} }}{{\mathop {\lim }\limits_{n \to \infty } \sum\limits_{i = 1}^{n - 1} {{p_{{\rm{ap}}}}(i)} }}\\
 &= p\left( {{x_m}} \right)
\end{split}
\end{equation}

In a SPAD array receiver, each pixel detects the optical signal once during a symbol duration. The avalanche probability depends on all possible symbol information. Hence, $p_{\rm{a}}^{\rm{A}}$ is given by
\begin{equation}\label{eq11}
p_{\rm{a}}^{\rm{A}} = \frac{{\mathop {\lim }\limits_{n \to \infty } \sum\limits_{i = 1}^{n - 1} {p\left( {{x_{m(n - i)}}} \right){p_{{\rm{ap}}}}(i)} }}{{\mathop {\lim }\limits_{n \to \infty } \sum\limits_{i = 1}^{n - 1} {{p_{{\rm{ap}}}}(i)} }}
\end{equation}

Assuming that all symbols are independent and the probability of transmitting an arbitrary symbol is equal to ${\textstyle{\frac{1}{M}}}$, where $M$ is the modulation order, the avalanche probabilities of every gate are mutually independent and identically distributed. The first-order moment estimation of $p_{\rm{a}}^{\rm{A}}$ is simplified to
\begin{equation}\label{eq12}
\begin{split}
p_{\rm{a}}^{\rm{A}} &\cong \frac{{\mathop {\lim }\limits_{n \to \infty } \sum\limits_{i = 1}^{n - 1} {\left\{ {\left( {\sum\limits_{m = 1}^M {\frac{{p\left( {{x_m}} \right)}}{M}} } \right){p_{{\rm{ap}}}}(i)} \right\}} }}{{\mathop {\lim }\limits_{n \to \infty } \sum\limits_{i = 1}^{n - 1} {{p_{{\rm{ap}}}}(i)} }}\\
 &= \frac{{\sum\limits_{m = 1}^M {\frac{{p\left( {{x_m}} \right)}}{M}} \mathop {\lim }\limits_{n \to \infty } \sum\limits_{i = 1}^{n - 1} {{p_{{\rm{ap}}}}(i)} }}{{\mathop {\lim }\limits_{n \to \infty } \sum\limits_{i = 1}^{n - 1} {{p_{{\rm{ap}}}}(i)} }}\\
 &= \sum\limits_{m = 1}^M {\frac{{p\left( {{x_m}} \right)}}{M}} 
\end{split}
\end{equation}

In conclusion, let $P\left( {{x_m}} \right) \buildrel \Delta \over = {P^{\rm{S}}}\left( {{x_m}} \right)$ denote the asymptotic trigger probability of a single SPAD receiver, and $P\left( {{x_m}} \right) \buildrel \Delta \over = {P^{\rm{A}}}\left( {{x_m}} \right)$ denote the asymptotic trigger probability of a SPAD array receiver. The expressions are written as follows:
\begin{equation}\label{eq13}
{P^{\rm{S}}}\left( {{x_m}} \right) = p\left( {{x_m}} \right) + Cp\left( {{x_m}} \right)\left( {1 - p\left( {{x_m}} \right)} \right)
\end{equation}
\begin{equation}\label{eq14}
 {P^{\rm{A}}}\left( {{x_m}} \right) = p\left( {{x_m}} \right) + C\sum\limits_{m = 1}^M {\frac{{p\left( {{x_m}} \right)}}{M}} \left( {1 - p\left( {{x_m}} \right)} \right)
\end{equation}

To obtain a simplified closed-form PMF of photon counts during a symbol duration, we study the asymptotic trigger probability. In our previous research \cite{ref43}, it has been proved that the detected photon counts during a symbol duration follow a binomial distribution $\mathsf{\mathbb{B}}\left( N,P\left( {{x}_{m}} \right) \right)$. Thus, we define the probability of detecting $k$  photon counts, denoted as ${P^\mathsf{\mathbb{B}}}\left(  \cdot  \right)$, during the symbol duration of  $''{{x}_{m}}''$. The PMF of photon counts is derived as follows:
\begin{equation} \label{eq15}
{P^\mathsf{\mathbb{B}}}\left( k \right) = \binom{N}{k}P{\left( {{x_m}} \right)^k}{\left( {1 - P\left( {{x_m}} \right)} \right)^{N - k}}
\end{equation}
where $N$ is the number of gates during the symbol duration, and $\binom{N}{k}$ is the combination number. It should be noted that $P\left( {{x_m}} \right) \buildrel \Delta \over = {P^{\rm{S}}}\left( {{x_m}} \right)$ for a single SPAD receiver, and $P\left( {{x_m}} \right) \buildrel \Delta \over = {P^{\rm{A}}}\left( {{x_m}} \right)$ for a SPAD array receiver.
\subsection{PMF Comparisons}

\begin{figure*}[!t]
\centering
\subfloat[]{\includegraphics[width=3.5in]{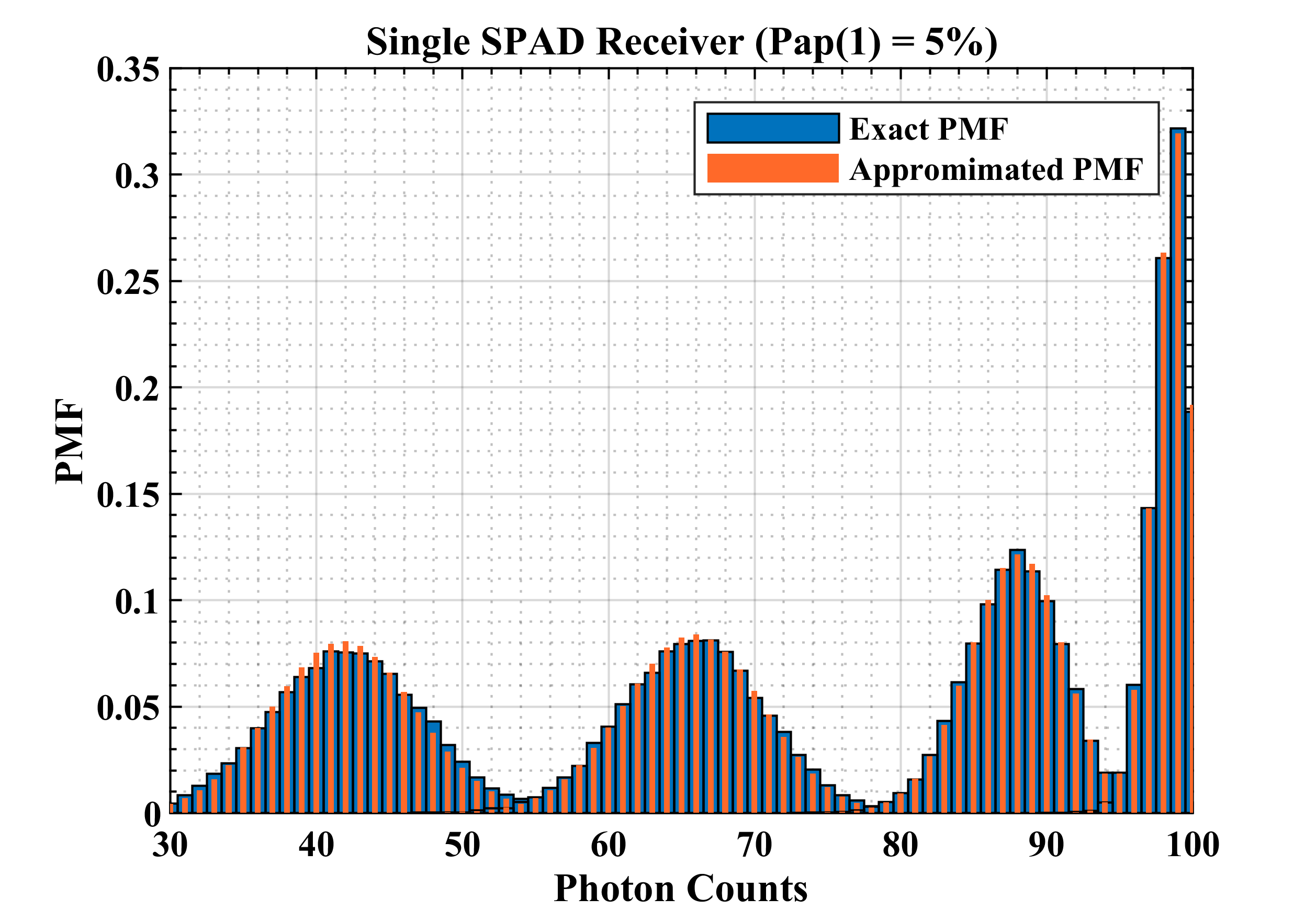}%
\label{fig_5_1}}
\hfil
\subfloat[]{\includegraphics[width=3.5in]{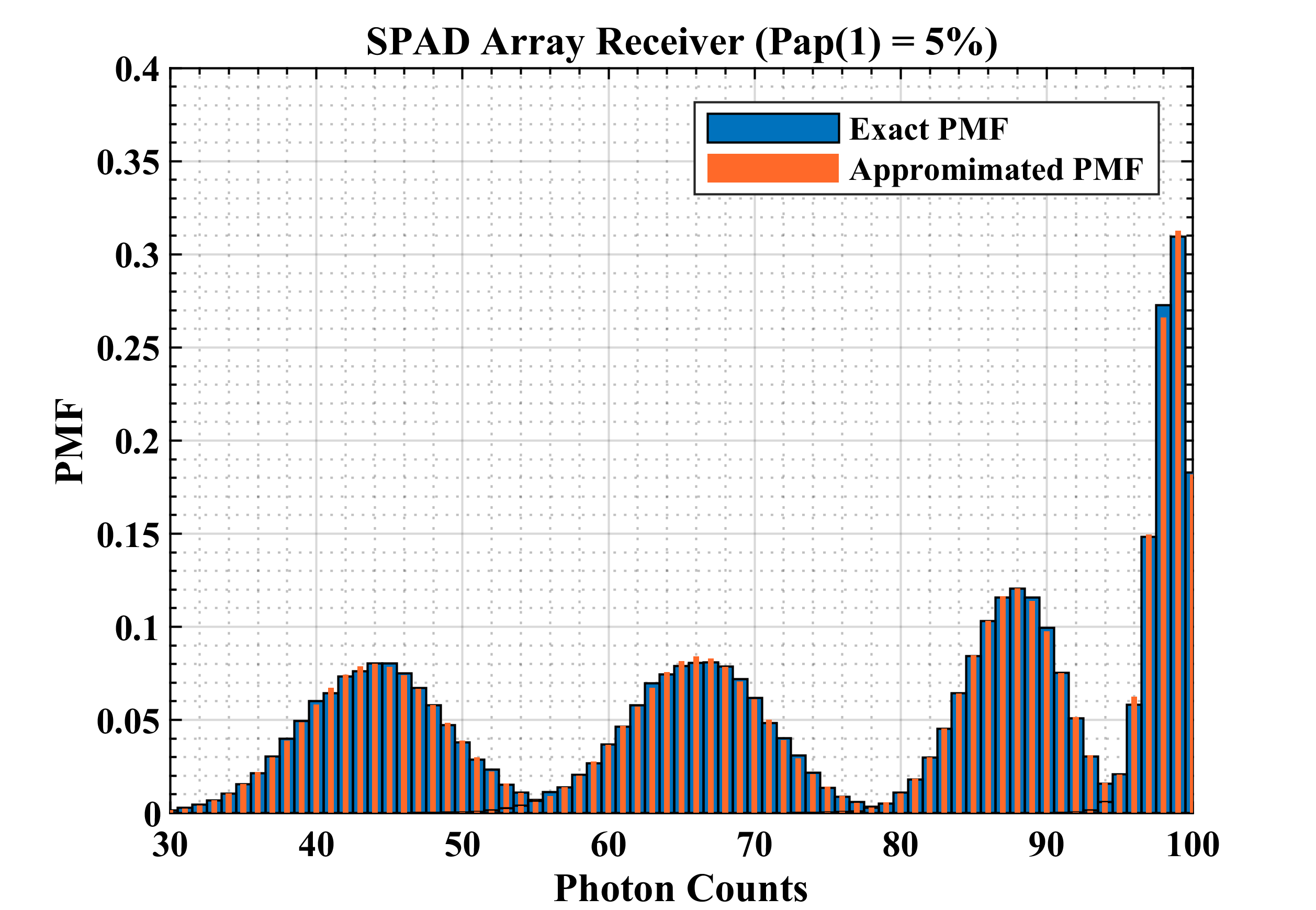}%
\label{fig_5_2}}
\hfil
\subfloat[]{\includegraphics[width=3.5in]{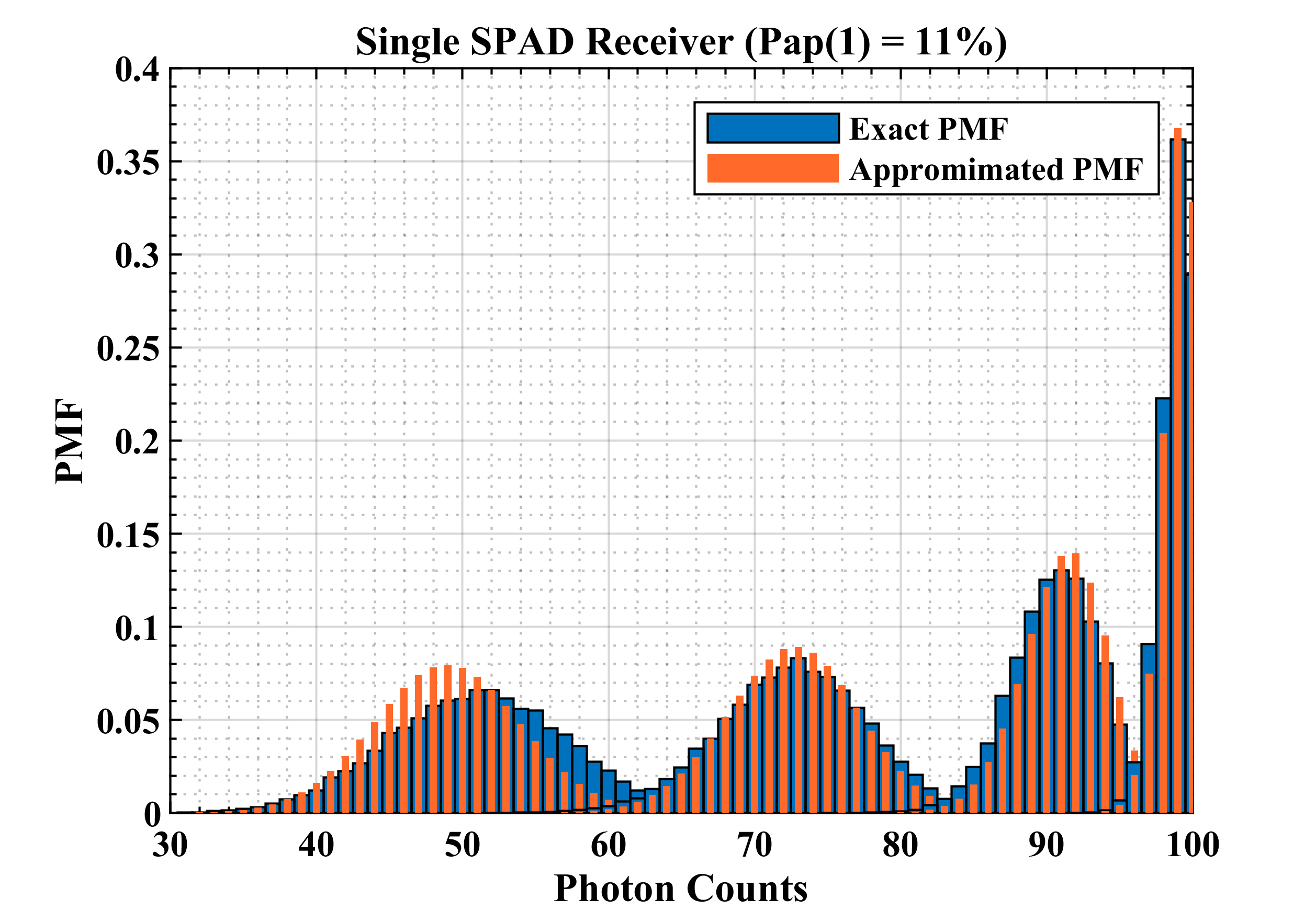}%
\label{fig_5_3}}
\hfil
\subfloat[]{\includegraphics[width=3.5in]{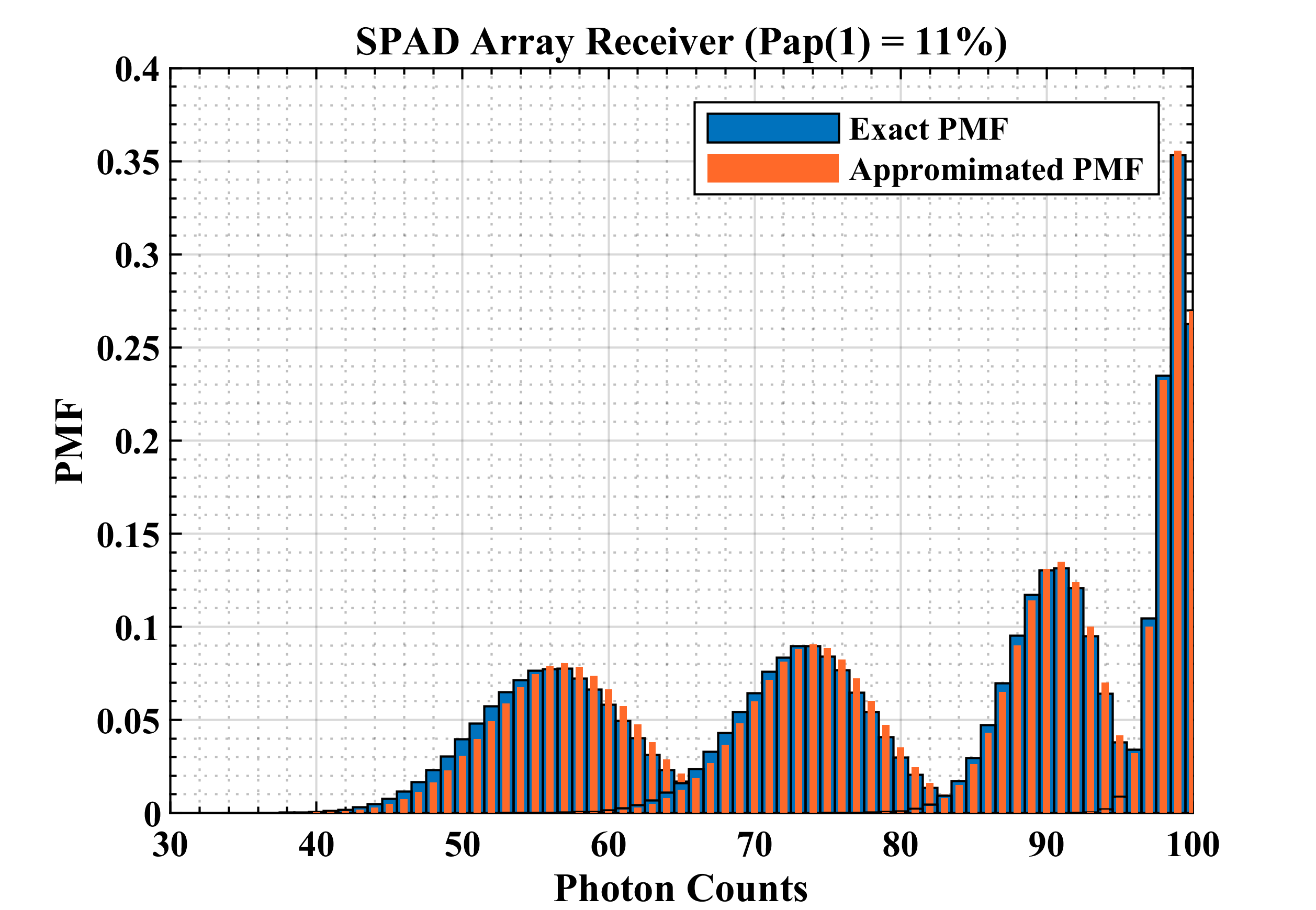}%
\label{fig_5_4}}
\caption{The exact and approximate PMFs of the detected photon counts for single SPAD and SAPD array based photon-counting receivers (${{\lambda }_{\text{s}}}=1,2,4,8{\text{c}}/{\text{ns}}\;,N=100$).}
\label{fig_5}
\end{figure*}

It is worth noting that the asymptotic expressions for the trigger probability are approximations, as the exact trigger probability for any given symbol duration depends on the sequences of all previously transmitted symbols. In theory, the accurate trigger probability should be calculated by incorporating all previous symbol information into (\ref{eq5}). However, the PMFs of photon counts are not identical for different symbols, even if they have the same symbol information. Using this approach leads to an exponential increase in computational complexity, by a factor of ${M^l}$, where $l$ represents the length of the past symbols sequence. For this reason, and to simplify the calculation, we introduce the asymptotic trigger probability as a means to characterize the photon-counting statistics.

Fig. 5 illustrates the discrepancy between the exact and the approximate PMFs of detected photon counts for 4-PAM. In reality, the asymptotic trigger probability represents a first-order moment estimation of the PMFs for different symbols, resulting in a mismatch between the exact and the approximate PMFs. This discrepancy is evident in Fig. 5c and 5d at a higher first-order AP of $11{\rm{\% }}$, where the tails of the exact PMFs are slightly lower or higher than those of the approximate PMFs. On the other hand, as shown in Fig. 5a and 5b, the contour profiles of the approximate and exact PMFs align remarkably well with each other at a lower first-order AP of  $5{\rm{\% }}$. These findings suggest that the analytical approximate PMF is capable of predicting the counting statistics for both single SPAD and SPAD array based photon-counting receivers, even when considering the afterpulsing effect.

\section{Signal Estimation and Decision Method}
\subsection{General Scheme}

\newcounter{TempEqCnt} 
\setcounter{TempEqCnt}{\value{equation}} 
\setcounter{equation}{15} 
\begin{figure*}[ht] 
\begin{align}\label{eq16}
SER = \frac{1}{M}\sum\limits_{m = 1}^{M} {\left\{ {\sum\limits_{k = 0}^{\left\lfloor {kth\left( {{x_m}} \right)} \right\rfloor } {\binom{N}{k}P{{\left( {{x_{m + 1}}} \right)}^k}} } \right.} 
 {\left( {1 - P\left( {{x_{m + 1}}} \right)} \right)^{N - k}}
 \left. { + \sum\limits_{k = \left\lceil {kth\left( {{x_m}} \right)} \right\rceil}^N {\binom{N}{k}P{{\left( {{x_m}} \right)}^k}{{\left( {1 - P\left( {{x_m}} \right)} \right)}^{N - k}}} } \right\}     
\end{align}
\hrulefill  
\end{figure*}

Using the approximate PMF of detected photon counts, we derive the expression for the average SER as shown in (\ref{eq16}), where $\left\lfloor  \cdot  \right\rfloor $ and $\left\lceil  \cdot  \right\rceil$ denote the floor and ceiling functions, respectively. Based on the maximum likelihood (ML) criterion, we derive a closed-form expression for the decision threshold.
\begin{align}\label{eq17}
kth\left( {{x_m}} \right) = \frac{{N\ln \frac{{1 - P\left( {{x_m}} \right)}}{{1 - P\left( {{x_{m + 1}}} \right)}}}}{{\ln \frac{{P\left( {{x_{m + 1}}} \right)\left( {1 - P\left( {{x_m}} \right)} \right)}}{{P\left( {{x_m}} \right)\left( {1 - P\left( {{x_{m + 1}}} \right)} \right)}}}}
\end{align}

In a SPAD receiver, the received optical intensity is estimated from the photon counts detected during the symbol duration. To mitigate the history-dependent afterpulsing effect, we introduce a retransmission scheme. For the pilot symbol sequence ${x_1},{x_2}, \ldots ,{x_M}$, each symbol is retransmitted $L$ times. We then calculate the mean of the detected photon counts for all symbol durations corresponding to the same symbol information. Using (\ref{eq7}), the estimation of avalanche probability is derived as follows:
\begin{equation}\label{eq18}
\hat p\left( {{x_m}} \right) = \frac{{\sqrt {{C^2} + 4P'\left( {{x_m}} \right)\left( {1 - C} \right)}  - C}}{{2\left( {1 - C} \right)}}
\end{equation}
with
\begin{equation}\label{eq19}
P'\left( {{x_m}} \right) = \frac{{\bar X}}{N}
\end{equation}
where $\bar X$ is the mean value of detected photon counts for all  symbol durations labeled with the same symbol $''{{x}_{m}}''$. Next, using (\ref{eq7}), (\ref{eq11}), and (\ref{eq13}), the estimation of trigger probability is given by
\begin{equation}\label{eq20}
\hat{P}\left( {{x}_{m}} \right)=\hat{p}\left( {{x}_{m}} \right)+C{{\hat{p}}_{\text{a}}}\left( 1-\hat{p}\left( {{x}_{m}} \right) \right)
\end{equation}
where ${{\hat{p}}_{\text{a}}}=\hat{p}\left( {{x}_{m}} \right)$  for a single SPAD receiver, and ${{\hat{p}}_{\text{a}}}=\sum\limits_{m=1}^{M}{\frac{\hat{p}\left( {{x}_{m}} \right)}{M}}$  for a SPAD array receiver.

By substituting (\ref{eq18}) into (\ref{eq17}), the practical threshold expression is obtained as
\begin{equation}\label{eq21}
\hat{k}th\left( {{x}_{m}} \right)=\frac{N\ln \frac{1-\hat{P}\left( {{x}_{m}} \right)}{1-\hat{P}\left( {{x}_{m+1}} \right)}}{\ln \frac{\hat{P}\left( {{x}_{m+1}} \right)\left( 1-\hat{P}\left( {{x}_{m}} \right) \right)}{\hat{P}\left( {{x}_{m}} \right)\left( 1-\hat{P}\left( {{x}_{m+1}} \right) \right)}}
\end{equation}

The symbol information is determined by comparing the detected photon counts $X$ with the practical threshold, and the detailed process is presented in Algorithm 1.

\begin{algorithm}[H]
\caption{Signal Estimation and Decision}\label{alg:alg1}
\begin{algorithmic}
\STATE 
\STATE {\textbf{Input:} Detected photon counts} $X$
\STATE {\textbf{Output:} Demodulated symbol information} $Symbol$
\STATE \textbf{1:} Compute ${P}'\left( {{x}_{m}} \right)$ using (\ref{eq19}) and $\hat{p}\left( {{x}_{m}} \right)$ using (\ref{eq18})
\STATE \textbf{2:} Compute $\hat{P}\left( {{x}_{m}} \right)$ using (\ref{eq20}) and $\hat kth \left( {{x_m}}\right)$ using (\ref{eq21})
\STATE \textbf{3: If} $X<\hat{k}th\left( {{x}_{1}} \right)$ \textbf{then} ${Symbol}={{x}_{1}}$
\STATE \textbf{4: }\hspace{0.3cm} \textbf{Else if} $
\hat kth\left( {{x_{m - 1}}} \right) \le X < \hat kth\left( {{x_m}} \right)$  \textbf{then} \STATE \hspace{2.3cm}  ${Symbol} = {x_m}$
\STATE \textbf{5: }\hspace{0.7cm} \textbf{Else if} $
X \ge \hat kth\left( {{x_{M - 1}}} \right)$  \textbf{then} ${Symbol} = {x_M}$
\STATE \textbf{6: End if}
\STATE \textbf{7: Return} 
\end{algorithmic}
\label{alg1}
\end{algorithm}
The general scheme is suitable for both single SPAD and SPAD array based photon-counting receivers.

\subsection{History-Enhanced Scheme}
In a SPAD array, optical signals are detected $N$  times within a detection cycle. Leveraging this, we propose a history-enhanced signal estimation scheme. In this scheme, the symbols ${x_1},{x_2}, \ldots ,{x_M}$ need to be retransmitted only ${\textstyle{\frac{1}{M}}}$ times compared to the general scheme, substantially reducing the length of the pilot symbol sequence without compromising the accuracy of signal estimation. We use the detected photon counts from the last period of the pilot symbol sequence, and apply the trigger probability expression to solve quadratic equations with $M$ variables, which represent the correlations between adjacent gates:
\begin{align}\label{eq22}
\left\{ \begin{array}{l}
P'\left( {{x_1}} \right) = \hat p\left( {{x_1}} \right) + \left( {1 - \hat p\left( {{x_1}} \right)} \right)\left[ {\hat p\left( {{x_M}} \right){p_{{\rm{ap}}}}(1)} \right.\\
\left. { + \hat p\left( {{x_{M - 1}}} \right){p_{{\rm{ap}}}}(2) + ... + \hat p\left( {{x_2}} \right){p_{{\rm{ap}}}}(M - 1) + Cp_{\rm{a}}^{\rm{A}}} \right]\\
P'\left( {{x_2}} \right) = \hat p\left( {{x_2}} \right) + \left( {1 - \hat p\left( {{x_2}} \right)} \right)\left[ {\hat p\left( {{x_1}} \right){p_{{\rm{ap}}}}(1)} \right.\\
 + \left. {\hat p\left( {{x_M}} \right){p_{{\rm{ap}}}}(2) + ... + \hat p\left( {{x_3}} \right){p_{{\rm{ap}}}}(M - 1) + Cp_{\rm{a}}^{\rm{A}}} \right]\\
\begin{array}{*{20}{c}}
 \vdots & \vdots & \vdots & \vdots & \vdots & \vdots & \vdots & \vdots & \vdots & \vdots 
\end{array}\\
P'\left( {{x_M}} \right) = \hat p\left( {{x_M}} \right) + \left( {1 - \hat p\left( {{x_M}} \right)} \right)\left[ {\hat p\left( {{x_{M - 1}}} \right){p_{{\rm{ap}}}}(1)} \right.\\
\left. { + \hat p\left( {{x_{M - 2}}} \right){p_{{\rm{ap}}}}(2) + ... + \hat p\left( {{x_2}} \right){p_{{\rm{ap}}}}(M - 1) + Cp_{\rm{a}}^{\rm{A}}} \right]
\end{array} \right.
\end{align}
We employ the Newton-Raphson iteration method to solve these quadratic equations, obtaining the estimated avalanche probabilities which adhere to $
0 < \hat p\left( {{x_1}} \right) < \hat p\left( {{x_2}} \right) <  \ldots  < \hat p\left( {{x_M}} \right) < 1$. The symbol information is then demodulated using the signal decision algorithm. 
\subsection{SER Comparisons}

\begin{figure*}[!t]
\centering
\subfloat[]{\includegraphics[width=3.5in]{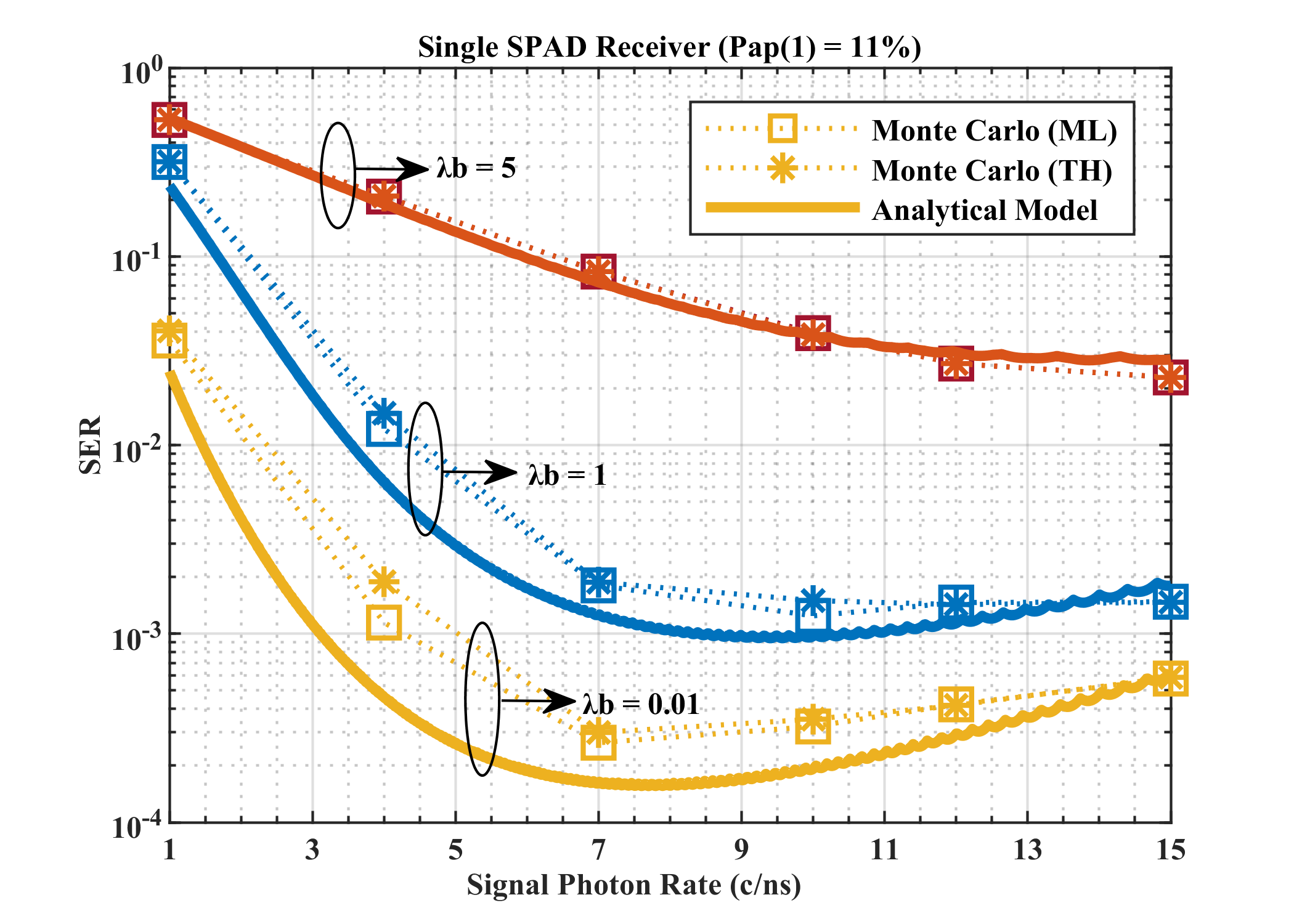}%
\label{fig_6_1}}
\hfil
\subfloat[]{\includegraphics[width=3.5in]{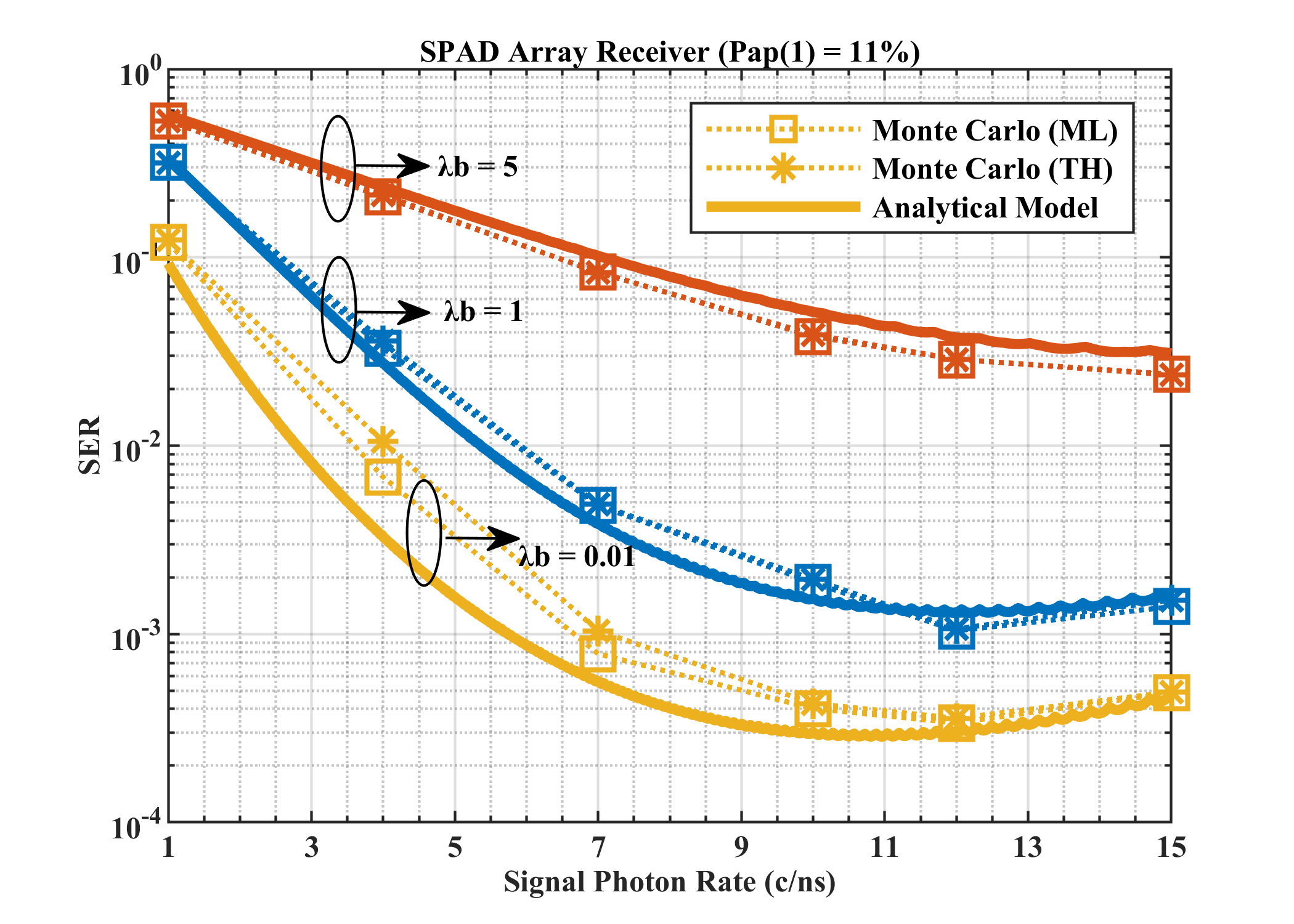}%
\label{fig_6_2}}
\caption{The exact and analytical SERs versus the received signal photon rate for single SPAD and SAPD array based photon-counting receivers ($N=256,{{p}_{\text{ap}}}\left( 1 \right)=11{\rm{\% }}$).}
\label{fig_6}
\end{figure*}

This section presents results of symbol error performance comparisons, where we juxtapose analytical findings with Monte Carlo simulation results. For our calculations and simulations, we assume independent statistics for each transmitted symbol, and consider the use of a high first-order AP ($
{p_{{\rm{ap}}}}\left( 1 \right) = 11{\rm{\% }}$), corresponding to a gate-ON interval (${\tau _{\rm{g}}} = 2{\rm{ns}}$) and detection cycle (${\tau _{{\rm{cys}}}} = 40{\rm{ns}}$). In all figures, we compare the analytically SER using (\ref{eq16}) with the Monte Carlo simulation results using ML detection and threshold (TH) detection from  (\ref{eq21}) respectively.

As showned in Fig. 6, we observe a remarkable agreement between ML and TH detection across all cases, confirming that within the defined parameter range, ML and TH detection are effectively equivalent. This validation underscores that the threshold defined by (\ref{eq21}) closely matches the ML threshold from the Monte Carlo simulations, emphasizing the significant improvement in error performance achieved using the threshold expressions. Furthermore, the analytical SER results align well with the Monte Carlo simulation outcomes, with discrepancies between the analytical model and simulations remaining within an order of magnitude below 1. Therefore, we can confidently affirm the validity of the analytical model for both single SPAD and SPAD array based photon-counting receivers.

\section{Numerical Results and Discussion}

\begin{table*}[!t]
\caption{The setting of simulation parameters\label{tab:table1}}
\centering
\begin{tabular}{|c|c|c|}
\hline
Notation & Description & Value\\
\hline
${p_{{\rm{de}}}}$ & Photon detection efficiency & $
10\% $\\
\hline
${\tau _{\rm{g}}}$ & Gate-ON interval& $1\sim10\text{ns}$\\
\hline
${{\tau }_{\text{d}}}$ & Dead time & $10\sim100\text{ns}$\\
\hline
${{\tau }_{\text{cyc}}}$ & Detection cycle & $50\sim100\text{ns}$ \\
\hline
${{\tau }_{\text{rel},j}}$ & Trap lifetime& $0.1,1.0,6.6,26.5,168.9,1078.7\text{ns}$ \\
\hline
${{A}_{j}}$ & Trap density& $
\text{4}\text{.95}\times \text{1}{{\text{0}}^{10}}\text{,4}\text{.70}\times \text{1}{{\text{0}}^{9}}\text{,6}\text{.72}\times \text{1}{{\text{0}}^{8}}\text{,1}\text{.54}\times \text{1}{{\text{0}}^{8}}\text{,2}\text{.08}\times \text{1}{{\text{0}}^{7}}\text{,2}\text{.53}\times \text{1}{{\text{0}}^{6}}\text{c}{{\text{m}}^{-3}}$\\
\hline
${{\lambda }_{\text{b}}}$ & Background photon rate & $
0.01\sim2{\text{c}}/{\text{ns}}\;$\\
\hline
${{\lambda }_{\text{d}}}$ & Dark carrier rate & $4.4\times {{10}^{-5}}{\text{c}}/{\text{ns}}\;$\\
\hline
${{N}_{\text{A}}}$ & Array scale& 256\\
\hline
\end{tabular}
\end{table*}

In this section, we investigate the symbol error performance in both single SPAD and SPAD array based photon-counting receivers. We utilize a 6-order exponential afterpulsing model, which has been previously described in our work on the InGaAs/InP SPAD QCD-300 \cite{ref44}. Since the QCD-300 was manufactured a decade ago, it suffers from a severe afterpulsing effect. To mitigate this effect, we set the dead time to be on the order of tens of nanoseconds, which is a relatively long interval. The background photon rate is measured to be ${\lambda _{\rm{b}}} = 0.01\sim2{{\rm{c}} \mathord{\left/
 {\vphantom {{\rm{c}} {{\rm{ns}}}}} \right.
 \kern-\nulldelimiterspace} {{\rm{ns}}}}$, as determined in a previous experiment \cite{ref43}. Additionally, we introduce the recent square-root signaling method to design the PAM signal constellation, which is well-suited for systems with signal-dependent noise \cite{ref54,ref55,ref56}. Building upon the foundation of square-root signaling, we make some modifications to the PAM signal constellation, resulting in final representations of $\left\{ {0,{\lambda _{\rm{s}}}} \right\}$ for 2-PAM, $\left\{ {0,0.25{\lambda _{\rm{s}}},0.56{\lambda _{\rm{s}}},{\lambda _{\rm{s}}}} \right\}$ for 4-PAM, and $
\left\{ {0,0.02{\lambda _{\rm{s}}},0.08{\lambda _{\rm{s}}},0.18{\lambda _{\rm{s}}},0.33{\lambda _{\rm{s}}},0.51{\lambda _{\rm{s}}},0.73{\lambda _{\rm{s}}},{\lambda _{\rm{s}}}} \right\}$ for 8-PAM, respectively. The simulation parameters are presented in Table \ref{tab:table1}.

\subsection{Single SPAD based Photon-Counting Receiver}

\begin{figure*}[!t]
\centering
\subfloat[]{\includegraphics[width=3.5in]{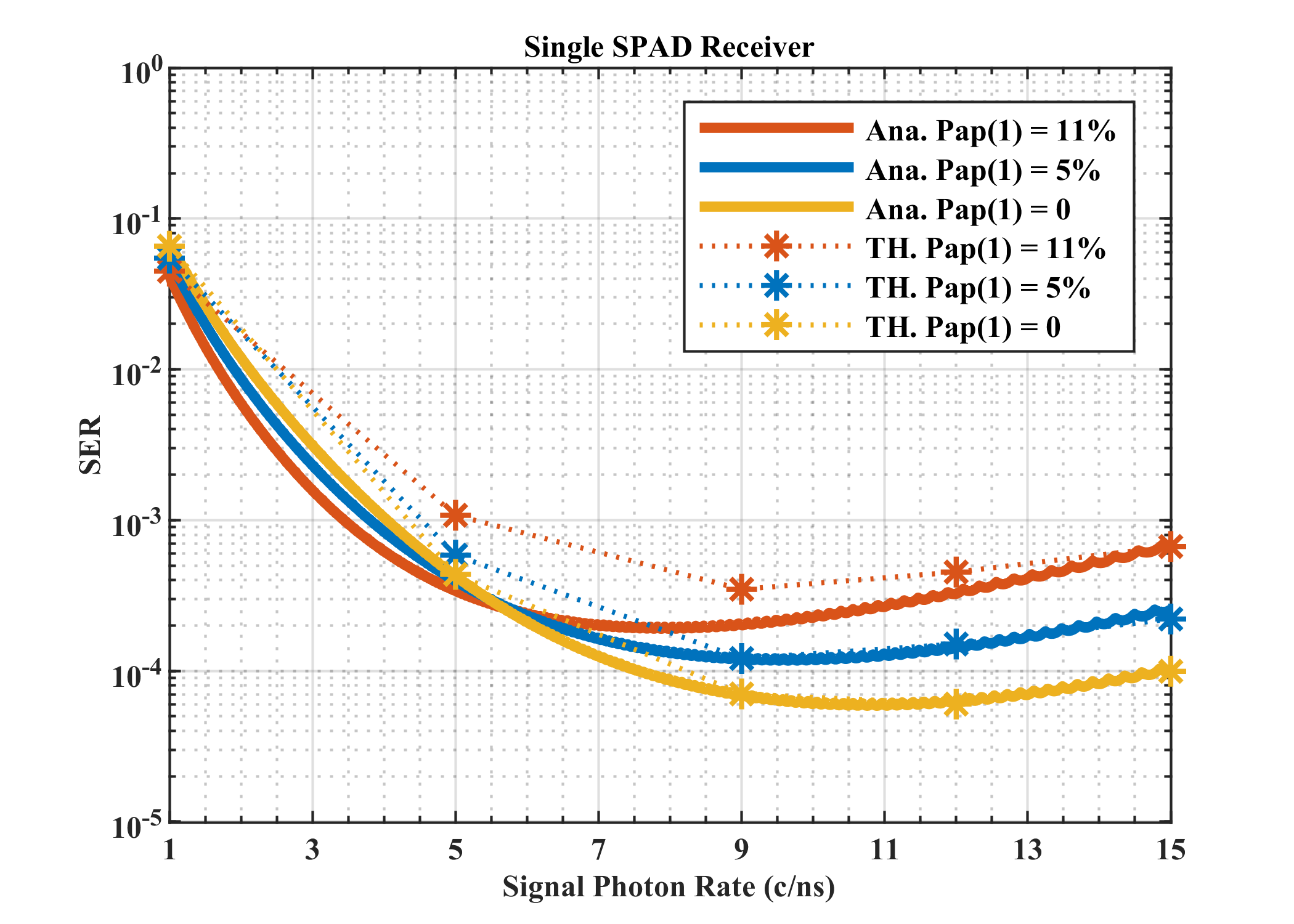}%
\label{fig_7_1}}
\hfil
\subfloat[]{\includegraphics[width=3.5in]{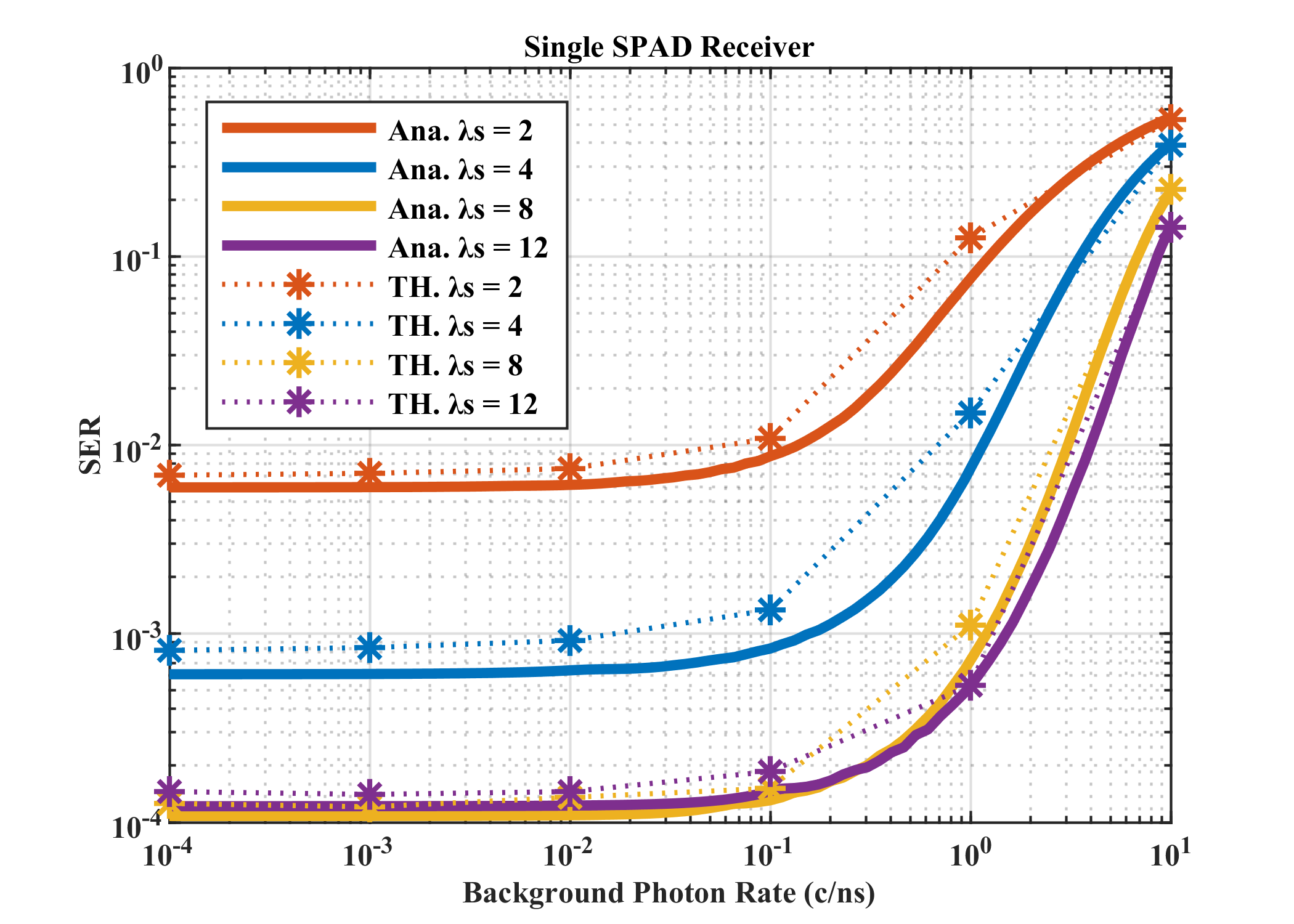}%
\label{fig_7_2}}
\caption{(a)The SER for a single SPAD based photon-counting receiver versus the received signal photon rate for 4-PAM under different APs (${{\lambda }_{\text{b}}}=0.1{\text{c}}/{\text{ns}}\;,N=256$). (b) The SER for a single SPAD based photon-counting receiver versus the received background photon rate for 4-PAM in different optical regimes ($N=256,{{p}_{\text{ap}}}\left( 1 \right)=5{\rm{\% }}$). }
\label{fig_7}
\end{figure*}

\begin{figure*}[!t]
\centering
\subfloat[]{\includegraphics[width=3.5in]{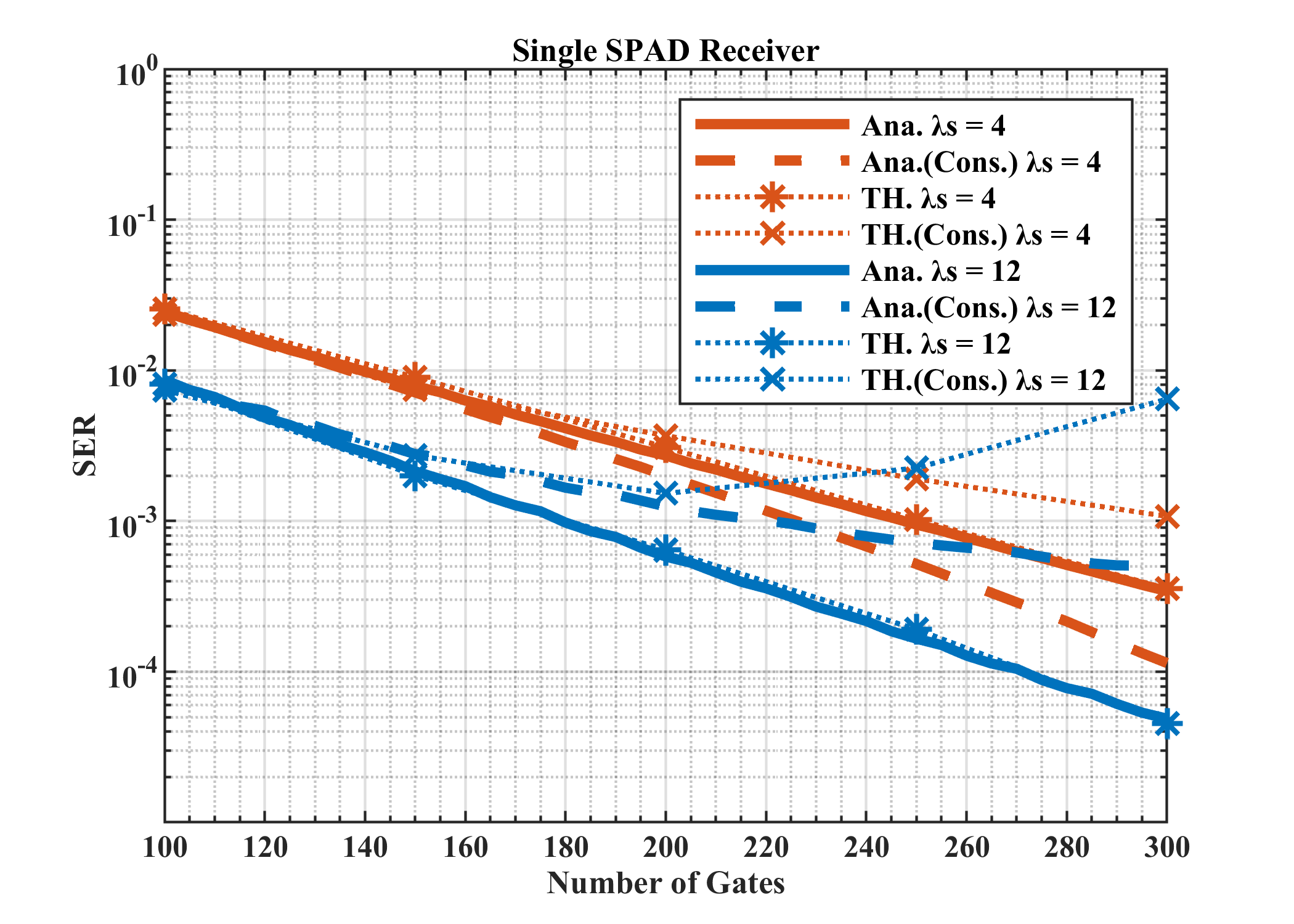}%
\label{fig_8_1}}
\hfil
\subfloat[]{\includegraphics[width=3.5in]{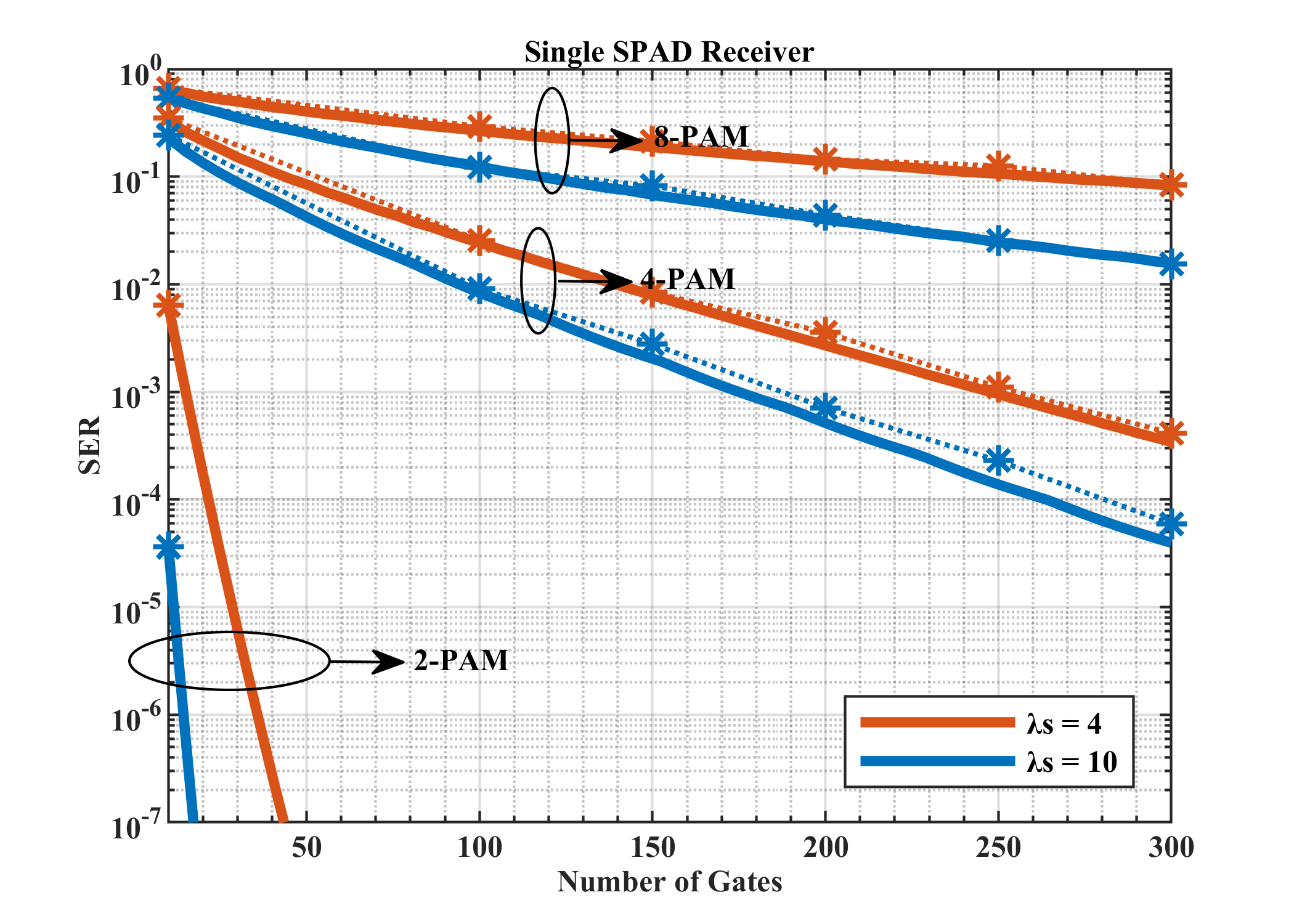}%
\label{fig_8_2}}
\caption{(a) The SER for a single SPAD based photon-counting receiver versus the number of gates for dynamic and constant symbol durations for 4-PAM in different optical regimes (${{\lambda }_{\text{b}}}=0.1{\text{c}}/{\text{ns}}\;,{{p}_{\text{ap}}}\left( 1 \right)=5{\rm{\% }}$). (b) The SER for a single SPAD based photon-counting receiver versus the number of gates for 2-PAM, 4-PAM and 8-PAM in different optical regimes (${{\lambda }_{\text{b}}}=0.1{\text{c}}/{\text{ns}}\;,{{p}_{\text{ap}}}\left( 1 \right)=5{\rm{\% }}$). }
\label{fig_8}
\end{figure*}

Fig. 7 and Fig. 8 present the SER results for a single SPAD based photon-counting receiver. In these figures, the symbol error probabilities for 4-PAM systems using TH detection are compared with Monte Carlo simulation results, resulting in perfectly matching curves for a first-order AP of $0 \sim 11\% $. The analytical calculations are based on the expressions provided in (\ref{eq13}), (\ref{eq16}), and (\ref{eq17}) for TH detection.

In Fig. 7a, assuming a moderately low background radiation of ${\lambda _{\rm{b}}} = 0.1{{\rm{c}} \mathord{\left/
 {\vphantom {{\rm{c}} {{\rm{ns}}}}} \right.
 \kern-\nulldelimiterspace} {{\rm{ns}}}}$, the SER initially decreases with a low signal photon rate and then increases when the signal photon rate exceeds ${{{\rm{8,10,11c}}} \mathord{\left/
 {\vphantom {{{\rm{8,10,11c}}} {{\rm{ns}}}}} \right.
 \kern-\nulldelimiterspace} {{\rm{ns}}}}$ for the corresponding first-order AP of $0,5\% ,11\% $ respectively. In low optical regimes, the improvement in the signal-to-noise ratio due to an increased signal photon rate dominates the SER improvement. However, when the signal photon rate surpasses a certain threshold, the detected photon counts become saturated, leading to a degradation in SER. Thus, maintaining an appropriate incident optical intensity is crucial. Moreover, a high AP value degrades system performance, as the SPAD becomes saturated with false photon counts generated by the afterpulsing effect.

In Fig. 7b, we investigate the effects of background radiation within the range of ${\lambda _{\rm{b}}} = {10^{ - 5}} \sim 10{{\rm{c}} \mathord{\left/{\vphantom {{\rm{c}} {{\rm{ns}}}}} \right.
 \kern-\nulldelimiterspace} {{\rm{ns}}}}$, which covers the majority of communication scenarios \cite{ref29}. As observed in Fig. 7b, the single SPAD receiver performance is highly dependent on  background radiation. The SERs rapidly degrade to ${10^{ - 1}}$ across all optical signal regimes of ${\lambda _{\rm{s}}} = 2,4,8,12{{\rm{c}} \mathord{\left/
 {\vphantom {{\rm{c}} {{\rm{ns}}}}} \right.\kern-\nulldelimiterspace} {{\rm{ns}}}}$, when the background photon rate exceeds  $1{{\rm{c}} \mathord{\left/{\vphantom {{\rm{c}} {{\rm{ns}}}}} \right. \kern-\nulldelimiterspace} {{\rm{ns}}}}$. This upper limit of background photon rate is known as the tolerated background radiation. On the other hand, the error probability floor is related to optical signal intensity, becoming more pronounced when $
{\lambda _{\rm{s}}}$ is below the appropriate value. It is evident that maintaining the background photon rate below the tolerated level is necessary to ensure optimal system performance, and achieving a desired SER requires a appropriate optical signal intensity.

Fig. 8a shows the SER results in relation to the number of gates, which can be achieved by either shortening the dead time or by extending the symbol duration. It is observed that the SER decreases exponentially with an increasing number of gates. When symbol duration is extended, the afterpulsing effect remains constant, and the SER gradients are identical across different optical signal regimes. However, when the number of gates is increased by shortening the dead time while keeping the symbol duration at ${T_{\rm{s}}} = 7500{\rm{ns}}$, the improvement in error performance gradually diminishes. As a result, the SER for ${\lambda _{\rm{s}}} = 12{{\rm{c}} \mathord{\left/{\vphantom {{\rm{c}} {{\rm{ns}}}}} \right.\kern-\nulldelimiterspace} {{\rm{ns}}}}$ is higher than that for ${\lambda _{\rm{s}}} = 4{{\rm{c}} \mathord{\left/
{\vphantom {{\rm{c}} {{\rm{ns}}}}} \right.\kern-\nulldelimiterspace} {{\rm{ns}}}}$. In this scenario, as the dead time decreases, the SPAD becomes more susceptible to a severe afterpulsing effect. When the afterpulsing effect ${p_{{\rm{ap}}}}\left( 1 \right) \ge 18\%$, TH detection results are higher than analytical results due to the analytical model assumption of ${p_{{\rm{ap}}}}\left( 1 \right) \le 20\% $. For the systems under consideration, increasing the number of gates is always an effective method for improving error performance. However, when shortening the dead time, the incident optical intensity must be carefully considered. Otherwise, the improvement in SER will be limited.

Fig. 8b investigates the gradients of SER with respect to the number of gates for different modulation orders. For 8-PAM, according to the gradient of curve, the number of gates $N = 600$ is needed to achieve a SER of ${10^{ - 3}}$ in the optical regime of ${\lambda _{\rm{s}}} = 8{{\rm{c}} \mathord{\left/
 {\vphantom {{\rm{c}} {{\rm{ns}}}}} \right.
 \kern-\nulldelimiterspace} {{\rm{ns}}}}$. For 4-PAM and 2-PAM, the system exhibits significantly better SERs and gradients. Due to the discrete distribution of photon counts, a larger number of gates is required to separate the PMFs of different symbol information and to improve error performance. It is worth noting that the maximum achievable modulation order depends on the number of gates.

\subsection{SPAD Array based Photon-Counting Receiver}

\begin{figure*}[!t]
\centering
\subfloat[]{\includegraphics[width=3.5in]{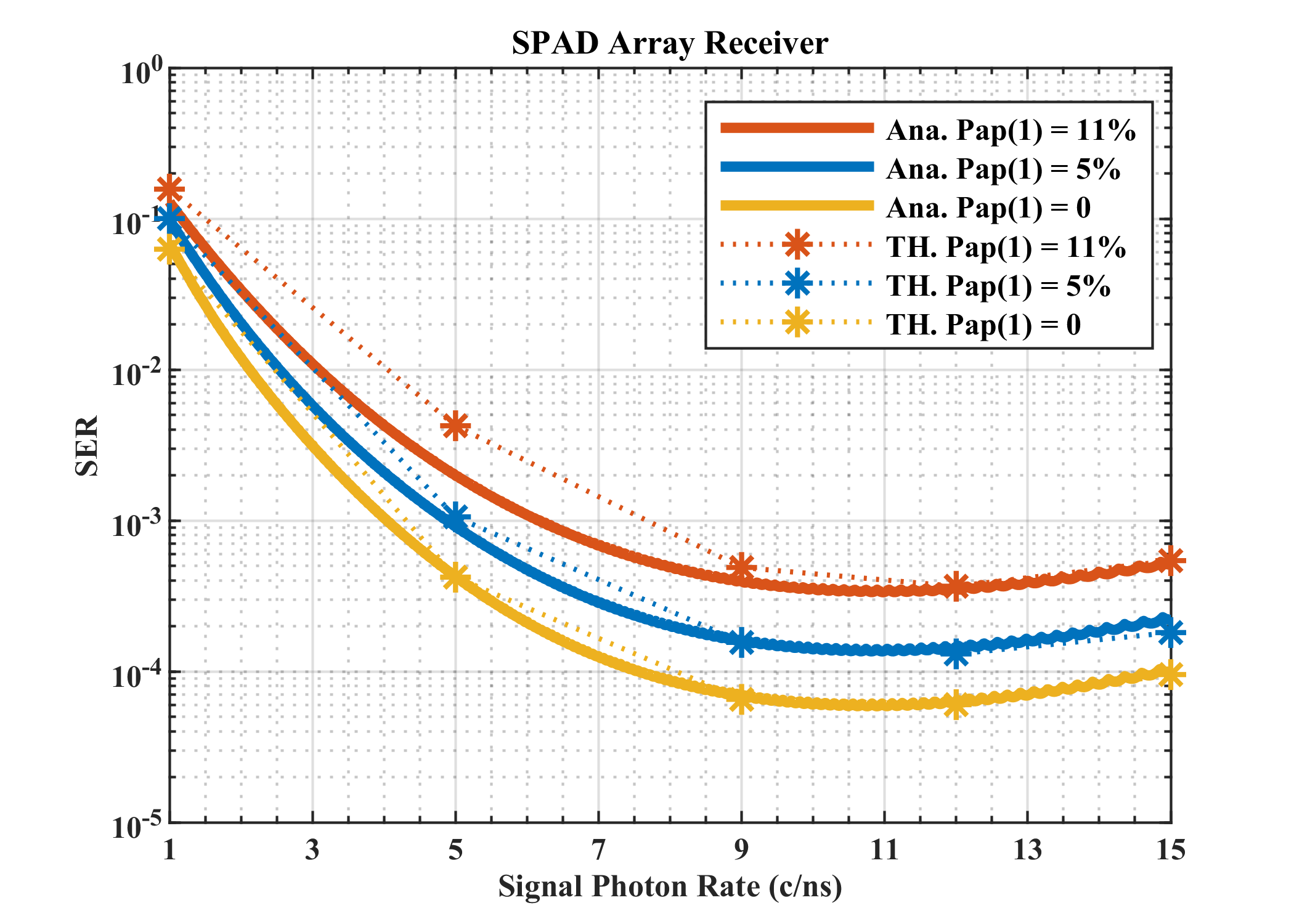}%
\label{fig_9_1}}
\hfil
\subfloat[]{\includegraphics[width=3.5in]{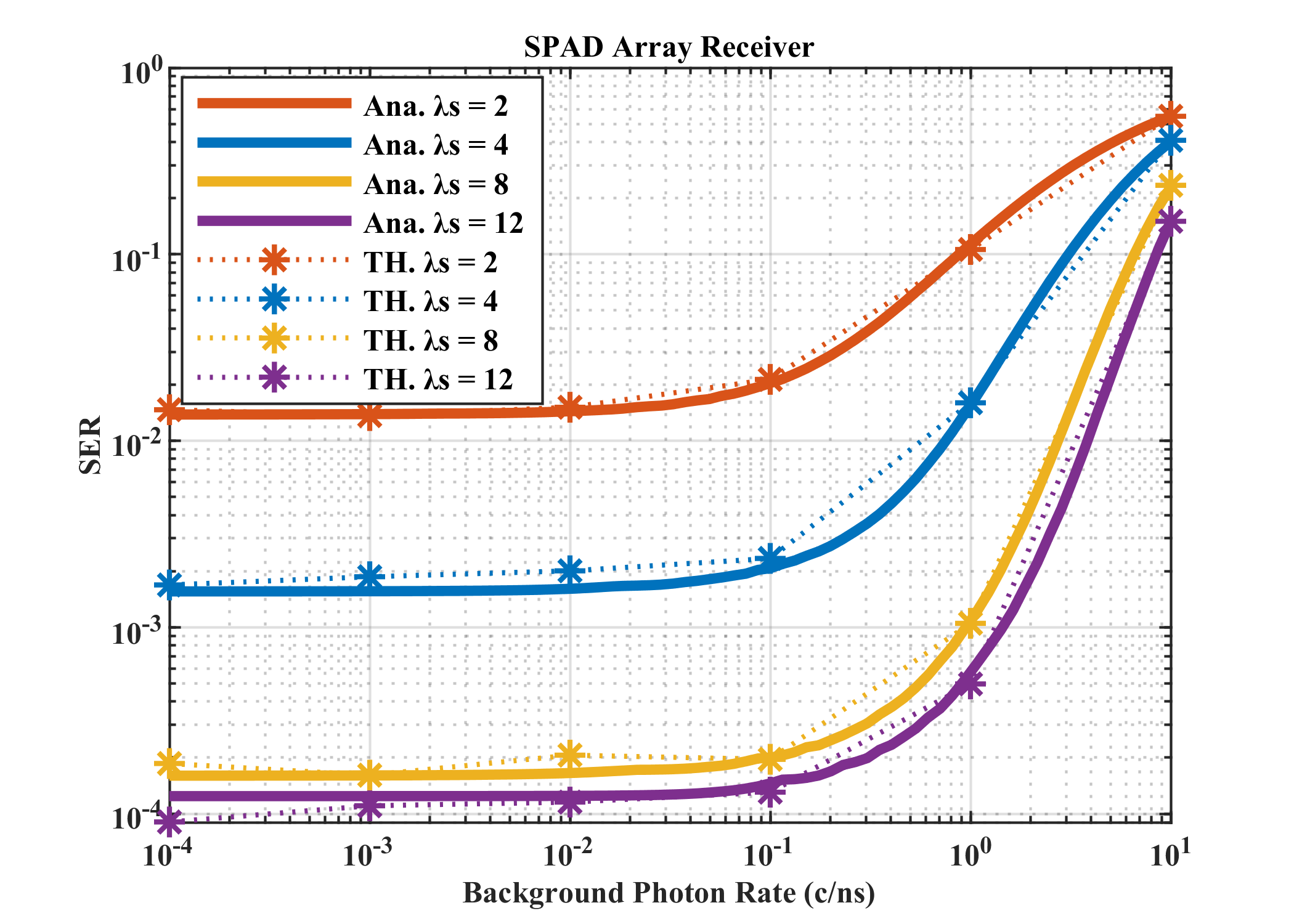}%
\label{fig_9_2}}
\caption{(a) The SER for a SPAD array based photon-counting receiver versus the received signal photon rate for 4-PAM under different APs (${{\lambda }_{\text{b}}}=0.1{\text{c}}/{\text{ns}}\;,N=256$). (b) The SER for a SPAD array based photon-counting receiver versus the received background photon rate for 4-PAM in different optical regimes  ($N=256,{{p}_{\text{ap}}}\left( 1 \right)=5{\rm{\% }}$). }
\label{fig_9}
\end{figure*}

\begin{figure*}[!t]
\centering
\subfloat[]{\includegraphics[width=3.5in]{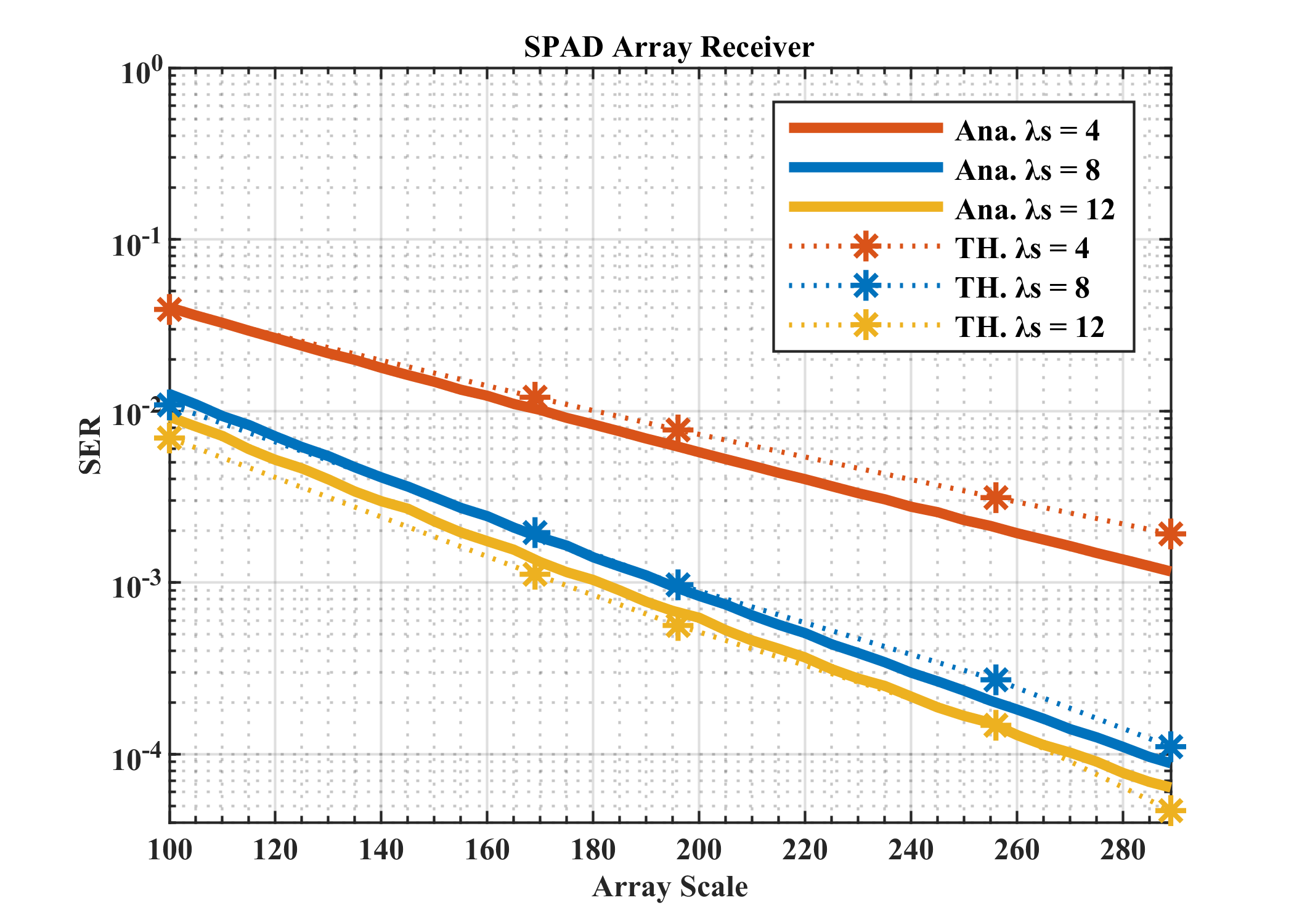}%
\label{fig_10_1}}
\hfil
\subfloat[]{\includegraphics[width=3.5in]{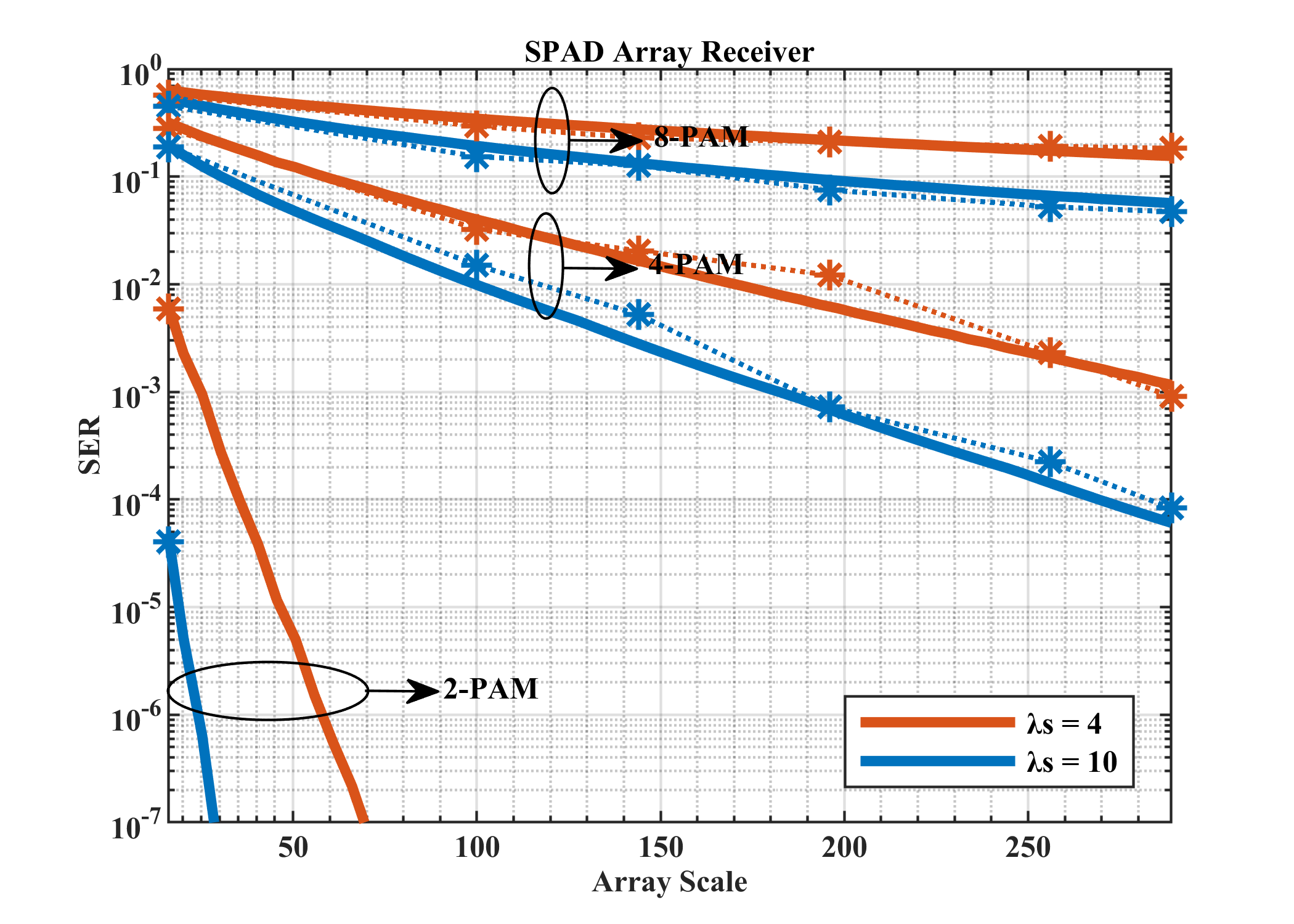}%
\label{fig_10_2}}
\caption{ (a) The SER for a SPAD array based photon-counting receiver versus the number of gates for 4-PAM in different optical regimes (${{\lambda }_{\text{b}}}=0.1{\text{c}}/{\text{ns}}\;,{{p}_{\text{ap}}}\left( 1 \right)=5{\rm{\% }}$). (b) The SER for a SPAD array based photon-counting receiver versus the number of gates for 2-PAM, 4-PAM and 8-PAM in different optical regimes  (${{\lambda }_{\text{b}}}=0.1{\text{c}}/{\text{ns}}\;,{{p}_{\text{ap}}}\left( 1 \right)=5{\rm{\% }}$). }
\label{fig_10}
\end{figure*}

Fig. 9 illustrates the error performance results for a SPAD array based photon-counting receiver. In this figure, we numerically evaluate the error probability with TH detection, as defined by (\ref{eq14}), (\ref{eq16}), and (\ref{eq17}), and compare it with Monte Caelo simulation results. The threshold value is numerically determined as well. Once again, we observe perfectly matching curves with the use of TH detection.

Similar to the SER results for the single SPAD receiver, three different values of AP are considered here. In Fig. 9a, we assume first-order AP values of $0,5\% ,11\%$. Once again, a high signal photon rate impairs the error performance and leads to SPAD saturation. Unlike the single SPAD, for a SPAD array, the optimal levels of incident optical signal intensity are identical at ${{{\rm{11c}}} \mathord{\left/
 {\vphantom {{{\rm{11c}}} {{\rm{ns}}}}} \right.
 \kern-\nulldelimiterspace} {{\rm{ns}}}}$ for different APs. Additionally, a higher AP value worsens the system error performance, as is evident in Fig. 9a.

As shown in Fig. 9b, the tolerated level of background radiation also plays a critical role in determining error performance. In scenarios where the background radiation level is similar to that of the SPAD array receiver, the SER floor is linked to optical signal intensity. Due to the stochastic nature of adjacent gates based on differing symbol information, the PMF of photon counts becomes broader, leading to SER degradation. Consequently, a more stringent limit on background radiation ${\lambda _{\rm{b}}} = {10^{ - 1}}{{\rm{c}} \mathord{\left/{\vphantom {{\rm{c}} {{\rm{ns}}}}} \right. \kern-\nulldelimiterspace} {{\rm{ns}}}}$ is introduced to maintain the target SER of a single SPAD. On the other hand, this limitation is necessary to achieve a higher information transfer rate for the SPAD array.

Fig. 10a presents the SER results concerning the number of gates for a SPAD array receiver. Similar to the single SPAD receiver, as the number of gates increases, the SER improves exponentially. For the SPAD array receiver, increasing the number of gates is achieved by expanding the array scale, which does not reduce the information transfer rate or increase the afterpulsing effect. Thus, increasing the number of gates remains the most effective method for improving error performance in SPAD array receivers.

As observed in Fig. 10b, the gradient of SER decreases with higher-order PAM. Due to the pronounced afterpulsing effect, the PMFs of photon counts become broader for SPAD array receivers. Consequently, for 8-PAM, the gradient of SER for the SPAD array is less steep than that for the single SPAD. Based on the curve profiles, a number of gates $N = 900$ should be provided to achieve a SER of ${10^{-3}}$. It is clear that the system margin for the SPAD array should be greater than that for the single SPAD.

\subsection{Error Performance Comparisons}
In summary, the error performance characteristics of SPAD array receivers are similar to those of single SPAD receivers. As mentioned in Section \uppercase\expandafter{\romannumeral3}, when using a single SPAD, most gates suffer from a higher-order afterpulse effect, and the correlations between adjacent gates are strong. Consequently, the amplification of photon counts due to the afterpulsing effect is significant in high optical regimes but negligible in low optical regimes. The PMF gaps between different optical signals can be further extended in single SPAD receivers compared to SPAD array receivers. Due to different afterpulsing effects on photon counts, this amplification for SPAD array receivers is relatively lower, and the saturation optical intensity is higher. Furthermore, the tolerated background radiation of SPAD array receivers is lower than that of single SPAD receivers. Based on the above analysis, maintaining a background radiation below the tolerance level and ensuring a certain number of gates are crucial for achieving a lower bound on the SER. Additionally, maintaining the signal intensity at an appropriate receiving intensity can further optimize the SER. Finally, the available modulation order is determined by the number of gates and the target SER.

\section{Conclusion}
In conclusion, we have presented a comprehensive analytical framework for modeling the statistical behavior of practical photon-counting receivers. Through our analysis, we have explored the impact of non-Markovian afterpulsing effect on photon-counting statistics, and formulated a closed-form approximation for the probability distribution of photon counts. Moreover, we have derived closed-form threshold expressions based on the ML criterion. We proposed a general signal estimation and decision scheme, as well as a specialized history-enhanced estimation scheme. Our statistical model accurately predicts the detected photon counts for high-order PAM signals in presence of dead time, non-photon-resolving and afterpulsing effect. By utilizing the proposed SER model, the parameters of FSO systems with practical photon-counting receivers can be designed and optimized efficiently. Overall, this study provides a reliable analytical model to evaluate the performance of photon-limited FSO systems with practical photon-counting receivers.

\section*{Appendix A \\Closed-Form Approximate Trigger Probability}\label{secA1}
In this appendix, we derive the closed-form approximation for the trigger probability. We begin by considering the trigger probabilities for the first, second, and third gates, while neglecting higher-order terms. This allows us to obtain the simplified expressions as follows: 
\begin{subequations}\label{eq23}
\begin{align}
{P_1} & = {p_1} \label{eq:23A}\\
{P_2} & = {p_2} + \left( {1 - {p_2}} \right){p_1}{p_{{\rm{ap}}}}(1)\label{eq:23B}\\
{P_3} & = {p_3} + \left( {1 - {p_3}} \right)\left[ {{p_1}{p_{{\rm{ap}}}}(2) + {p_2}{p_{{\rm{ap}}}}(1)} \right] \label{eq:23c}\\ 
{P_4} & = {p_4} + \left( {1 - {p_4}} \right)\left[ {{p_1}{p_{{\rm{ap}}}}(3) + {p_2}{p_{{\rm{ap}}}}(2) + {p_3}{p_{{\rm{ap}}}}(1)} \right] \label{eq:23d}
\end{align}
\end{subequations}

Building on this, we postulate that the closed-form approximation for ${P_{n - 1}}$ is as follows:
\begin{equation}\label{eq24}
{P_{n - 1}} = {p_{n - 1}} + \left( {1 - {p_{n - 1}}} \right)\sum\limits_{i = 1}^{n - 2} {{p_i}{p_{{\rm{ap}}}}(n - 1 - i)} 
\end{equation}
Progressing to the $n$-th gate, ${P_n}$ can be split into two components: the terms for the first gate where an avalanche event occurs (denoted as gate-1), represented by the first set $F\left( {{p_j},{p_{{\rm{ap}}}}(j)} \right)$, and the terms for first gate where no avalanche event occurs (denoted as gate-0), represented by the second set $G\left( {{p_j},{p_{{\rm{ap}}}}(j)} \right)$. The first set is further categorized into $i$-th gate-1 and $i$-th gate-0. The expression for ${P_n}$ is given as
\begin{align}\label{eq25}
\begin{array}{l}
{P_n} = F\left( {{p_j},{p_{{\rm{ap}}}}(j)} \right) + G\left( {{p_j},{p_{{\rm{ap}}}}(j)} \right)\\
 = {p_1}\left\{ {\sum\limits_{i = 2}^n {\left[ {{p_i} + \left( {1 - {p_i}} \right){p_{{\rm{ap}}}}(n - i)} \right]{a_i}\left( {{p_j},{p_{{\rm{ap}}}}(j)} \right)} } \right\}\\
 + {p_1}\left\{ {\sum\limits_{i = 2}^n {\left[ {1 - {p_i} - \left( {1 - {p_i}} \right){p_{{\rm{ap}}}}(n - i)} \right]{b_i}\left( {{p_j},{p_{{\rm{ap}}}}(j)} \right)} } \right\}\\
 + \left( {1 - {p_1}} \right)\left[ {{p_n} + \left( {1 - {p_n}} \right)\sum\limits_{i = 2}^{n - 1} {{p_i}{p_{{\rm{ap}}}}(n - i)} } \right]
\end{array}&
\end{align}
where ${a_i}\left( {{p_j},{p_{{\rm{ap}}}}(j)} \right)$ and $
{b_i}\left( {{p_j},{p_{{\rm{ap}}}}(j)} \right)$ are the remaining terms of $F\left( {{p_j},{p_{{\rm{ap}}}}(j)} \right)$  encompassing  ${p_{{\rm{ap}}}}(j)$, and ${p_j}$ signifies all terms of ${p_1},{p_2},...,{p_n}$. Next, we isolate all terms related to the afterpulse effect induced by the first gate-1, which includes terms like ${p_{{\rm{ap}}}}(n - 2)$ for second gate, ${p_{{\rm{ap}}}}(n - 3)$ for third gate, and so on. $
F\left( {{p_j},{p_{{\rm{ap}}}}(j)} \right)$  can be formulated as
\begin{align}\label{eq26}
\begin{array}{l}
F\left( {{p_j},{p_{{\rm{ap}}}}(j)} \right) = {p_1}\left\{ {\sum\limits_{i = 2}^n {\left( {1 - {p_i}} \right){p_{{\rm{ap}}}}(n - i){a_i}\left( {{p_j},{p_{{\rm{ap}}}}(j)} \right)} } \right\}\\
 - {p_1}\left\{ {\sum\limits_{i = 2}^n {\left( {1 - {p_i}} \right){p_{{\rm{ap}}}}(n - i){b_i}\left( {{p_j},{p_{{\rm{ap}}}}(j)} \right)} } \right\}\\
 + {p_1}\left[ {{p_n} + \left( {1 - {p_n}} \right)\sum\limits_{i = 2}^{n - 1} {{p_i}{p_{{\rm{ap}}}}(n - i)} } \right]
\end{array}
\end{align}
Disregarding the higher-order terms of AP, we abandon all terms with ${p_{{\rm{ap}}}}(j)$ in ${a_i}\left( {{p_j},{p_{{\rm{ap}}}}(j)} \right)$ and ${b_i}\left( {{p_j},{p_{{\rm{ap}}}}(j)} \right)$. For $i = 2 \sim n - 1$, ${a_i}\left( {{p_j},{p_{{\rm{ap}}}}(j)} \right)$ and ${b_i}\left( {{p_j},{p_{{\rm{ap}}}}(j)} \right)$ are approximated by the same expression as follows:
\begin{align}\label{eq27}
\begin{array}{c}
{a_i}\left( {{p_j},{p_{{\rm{ap}}}}(j)} \right) \cong {b_i}\left( {{p_j},{p_{{\rm{ap}}}}(j)} \right)\\
 \cong \left\{ {{p_2}{p_3}...{p_{i - 1}}{p_{i + 1}}...{p_n}} \right\} + \left\{ {(1 - {p_2}){p_3}...{p_{i - 1}}{p_{i + 1}}...{p_n}} \right\}\\
 + \left\{ {{p_2}(1 - {p_3})...{p_{i - 1}}{p_{i + 1}}...{p_n}} \right\} + ...\\
 + \left\{ {(1 - {p_2})(1 - {p_3})...(1 - {p_{i - 1}})(1 - {p_{i + 1}})...(1 - {p_n})} \right\}\\
 = 1
\end{array}
\end{align}
For $i = n$, there is only ${a_n}\left( {{p_j},{p_{{\rm{ap}}}}(j)} \right)$ without a corresponding ${b_n}\left( {{p_j},{p_{{\rm{ap}}}}(j)} \right)$. By substituting the approximate expressions of ${a_i}\left( {{p_j},{p_{{\rm{ap}}}}(j)} \right)$ and ${b_i}\left( {{p_j},{p_{{\rm{ap}}}}(j)} \right)$ into (\ref{eq25}), we derive the closed-form approximation for ${P_n}$ as
\begin{align}\label{eq28}
\begin{array}{c}
{P_n} = {p_n} + \left( {1 - {p_n}} \right)\sum\limits_{i = 2}^{n - 1} {{p_i}{p_{{\rm{ap}}}}(n - i)} \\
 + {p_1}\left( {1 - {p_n}} \right){p_{{\rm{ap}}}}(n - 1) \\
= {p_n} + \left( {1 - {p_n}} \right)\sum\limits_{i = 1}^{n - 1} {{p_i}{p_{{\rm{ap}}}}(n - i)} 
\end{array}
\end{align}
It is demonstrated that, based on the presupposed closed-form approximation for ${P_{n - 1}}$, ${P_n}$ is shown to conform to the same approximate expression. This completes the proof.

\section*{Appendix B \\Asymptotic Trigger Probability}\label{secB}
Using the closed-form approximation of trigger probability, the upper bound of trigger probability is computed as follows:
\begin{align}\label{eq29}
\begin{array}{c}
\mathop {\lim }\limits_{n \to \infty } {P_n} = {p_n} + \left( {1 - {p_n}} \right)\sum\limits_{i = 1}^{n - 1} {{p_i}{p_{{\rm{ap}}}}(n - i)} \\
 \le {p_n} + \max \left( {{p_1},{p_2},...,{p_n}} \right)\left( {1 - {p_n}} \right)\\
 \times \sum\limits_{i = 1}^{n - 1} {{p_{{\rm{ap}}}}(n - i)} \\
 = {p_n} + \max \left( {{p_1},{p_2},...,{p_n}} \right)\left( {1 - {p_n}} \right)\\
 \times \sum\limits_{i = 1}^{n - 1} {\left\{ {\int_{i{\tau _{{\rm{cyc}}}}}^{i{\tau _{{\rm{cyc}}}} + {\tau _{\rm{g}}}} {\left( {\sum\limits_j {{A_j}{{\rm{e}}^{ - {t \mathord{\left/
 {\vphantom {t {{\tau _{{\rm{rel,}}}}_j}}} \right.
 \kern-\nulldelimiterspace} {{\tau _{{\rm{rel,}}}}_j}}}}} } \right)} {\rm{d}}t} \right\}} \\
 = {p_n} + \max \left( {{p_1},{p_2},...,{p_n}} \right)\left( {1 - {p_n}} \right)\\
 \times \sum\limits_j {\left\{ {{A_j}\sum\limits_{i = 1}^{n - 1} {\left[ {\int_{i{\tau _{{\rm{cyc}}}}}^{i{\tau _{{\rm{cyc}}}} + {\tau _{\rm{g}}}} {{{\rm{e}}^{ - {t \mathord{\left/
 {\vphantom {t {{\tau _{{\rm{rel,}}}}_j}}} \right.
 \kern-\nulldelimiterspace} {{\tau _{{\rm{rel,}}}}_j}}}}} {\rm{d}}t} \right]} } \right\}} \\
 = {p_n} + \max \left( {{p_1},{p_2},...,{p_n}} \right)\left( {1 - {p_n}} \right)\\
 \times \sum\limits_j {\left\{ {{A_j}{\tau _{{\rm{rel,}}}}_j\sum\limits_{i = 1}^{n - 1} {\left[ {{{\rm{e}}^{ - {{i{\tau _{{\rm{cyc}}}}} \mathord{\left/
 {\vphantom {{i{\tau _{{\rm{cyc}}}}} {{\tau _{{\rm{rel,}}}}_j}}} \right.
 \kern-\nulldelimiterspace} {{\tau _{{\rm{rel,}}}}_j}}}}\left( {1 - {{\rm{e}}^{ - {{{\tau _{\rm{g}}}} \mathord{\left/
 {\vphantom {{{\tau _{\rm{g}}}} {{\tau _{{\rm{rel,}}}}_j}}} \right.
 \kern-\nulldelimiterspace} {{\tau _{{\rm{rel,}}}}_j}}}}} \right)} \right]} } \right\}} \\
 = {p_n} + \max \left( {{p_1},{p_2},...,{p_n}} \right)\left( {1 - {p_n}} \right)\\
 \times \sum\limits_j {\left\{ {{A_j}{\tau _{{\rm{rel,}}}}_j\frac{{{{\rm{e}}^{{{{\tau _{\rm{g}}}} \mathord{\left/
 {\vphantom {{{\tau _{\rm{g}}}} {{\tau _{{\rm{rel,}}}}_j}}} \right.
 \kern-\nulldelimiterspace} {{\tau _{{\rm{rel,}}}}_j}}}} - 1}}{{{{\rm{e}}^{{{{\tau _{\rm{g}}}} \mathord{\left/
 {\vphantom {{{\tau _{\rm{g}}}} {{\tau _{{\rm{rel,}}}}_j}}} \right.
 \kern-\nulldelimiterspace} {{\tau _{{\rm{rel,}}}}_j}}}}\left( {{{\rm{e}}^{{{{\tau _{{\rm{cyc}}}}} \mathord{\left/
 {\vphantom {{{\tau _{{\rm{cyc}}}}} {{\tau _{{\rm{rel,}}}}_j}}} \right.
 \kern-\nulldelimiterspace} {{\tau _{{\rm{rel,}}}}_j}}}} - 1} \right)}}} \right\}} 
\end{array}
\end{align}

Likewise, the lower bound of trigger probability is computed as follows:
\begin{align}\label{eq30}
\begin{array}{c}
\mathop {\lim }\limits_{n \to \infty } {P_n} \ge {p_n} + \min \left( {{p_1},{p_2},...,{p_n}} \right)\\
 \times \left( {1 - {p_n}} \right)\sum\limits_{i = 1}^{n - 1} {{p_{{\rm{ap}}}}(n - i)} \\
 = {p_n} + \min \left( {{p_1},{p_2},...,{p_n}} \right)\left( {1 - {p_n}} \right)\\
 \times \sum\limits_j {\left\{ {{A_j}{\tau _{{\rm{rel,}}}}_j\frac{{{{\rm{e}}^{{{{\tau _{\rm{g}}}} \mathord{\left/
 {\vphantom {{{\tau _{\rm{g}}}} {{\tau _{{\rm{rel,}}}}_j}}} \right.
 \kern-\nulldelimiterspace} {{\tau _{{\rm{rel,}}}}_j}}}} - 1}}{{{{\rm{e}}^{{{{\tau _{\rm{g}}}} \mathord{\left/
 {\vphantom {{{\tau _{\rm{g}}}} {{\tau _{{\rm{rel,}}}}_j}}} \right.
 \kern-\nulldelimiterspace} {{\tau _{{\rm{rel,}}}}_j}}}}\left( {{{\rm{e}}^{{{{\tau _{{\rm{cyc}}}}} \mathord{\left/
 {\vphantom {{{\tau _{{\rm{cyc}}}}} {{\tau _{{\rm{rel,}}}}_j}}} \right.
 \kern-\nulldelimiterspace} {{\tau _{{\rm{rel,}}}}_j}}}} - 1} \right)}}} \right\}} 
\end{array}
\end{align}

These can be summarized in the following manner:
\begin{align}\label{eq31}
\begin{array}{l}
{p_n} + \min \left( {{p_1},{p_2},...,{p_n}} \right)\\
 \times \left( {1 - {p_n}} \right)\sum\limits_j {\left\{ {{A_j}{\tau _{{\rm{rel,}}}}_j\frac{{{{\rm{e}}^{{{{\tau _{\rm{g}}}} \mathord{\left/
 {\vphantom {{{\tau _{\rm{g}}}} {{\tau _{{\rm{rel,}}}}_j}}} \right.
 \kern-\nulldelimiterspace} {{\tau _{{\rm{rel,}}}}_j}}}} - 1}}{{{{\rm{e}}^{{{{\tau _{\rm{g}}}} \mathord{\left/
 {\vphantom {{{\tau _{\rm{g}}}} {{\tau _{{\rm{rel,}}}}_j}}} \right.
 \kern-\nulldelimiterspace} {{\tau _{{\rm{rel,}}}}_j}}}}\left( {{{\rm{e}}^{{{{\tau _{{\rm{cyc}}}}} \mathord{\left/
 {\vphantom {{{\tau _{{\rm{cyc}}}}} {{\tau _{{\rm{rel,}}}}_j}}} \right.
 \kern-\nulldelimiterspace} {{\tau _{{\rm{rel,}}}}_j}}}} - 1} \right)}}} \right\}} \\
 \le \mathop {\lim }\limits_{n \to \infty } {P_n}\\
 \le {p_n} + \max \left( {{p_1},{p_2},...,{p_n}} \right)\left( {1 - {p_n}} \right)\\
 \times \sum\limits_j {\left\{ {{A_j}{\tau _{{\rm{rel,}}}}_j\frac{{{{\rm{e}}^{{{{\tau _{\rm{g}}}} \mathord{\left/
 {\vphantom {{{\tau _{\rm{g}}}} {{\tau _{{\rm{rel,}}}}_j}}} \right.
 \kern-\nulldelimiterspace} {{\tau _{{\rm{rel,}}}}_j}}}} - 1}}{{{{\rm{e}}^{{{{\tau _{\rm{g}}}} \mathord{\left/
 {\vphantom {{{\tau _{\rm{g}}}} {{\tau _{{\rm{rel,}}}}_j}}} \right.
 \kern-\nulldelimiterspace} {{\tau _{{\rm{rel,}}}}_j}}}}\left( {{{\rm{e}}^{{{{\tau _{{\rm{cyc}}}}} \mathord{\left/
 {\vphantom {{{\tau _{{\rm{cyc}}}}} {{\tau _{{\rm{rel,}}}}_j}}} \right.
 \kern-\nulldelimiterspace} {{\tau _{{\rm{rel,}}}}_j}}}} - 1} \right)}}} \right\}} 
\end{array}
\end{align}

Therefore, according to the intermediate value theorem, the asymptotic trigger probability is expressed as:
\begin{align}\label{eq33}
\begin{array}{l}
\mathop {\lim }\limits_{n \to \infty } {P_n} = {p_n} + {p_{\rm{a}}}\left( {1 - {p_n}} \right)\\
 \times \sum\limits_j {\left\{ {{A_j}{\tau _{{\rm{rel,}}}}_j\frac{{{{\rm{e}}^{{{{\tau _{\rm{g}}}} \mathord{\left/
 {\vphantom {{{\tau _{\rm{g}}}} {{\tau _{{\rm{rel,}}}}_j}}} \right.
 \kern-\nulldelimiterspace} {{\tau _{{\rm{rel,}}}}_j}}}} - 1}}{{{{\rm{e}}^{{{{\tau _{\rm{g}}}} \mathord{\left/
 {\vphantom {{{\tau _{\rm{g}}}} {{\tau _{{\rm{rel,}}}}_j}}} \right.
 \kern-\nulldelimiterspace} {{\tau _{{\rm{rel,}}}}_j}}}}\left( {{{\rm{e}}^{{{{\tau _{{\rm{cyc}}}}} \mathord{\left/
 {\vphantom {{{\tau _{{\rm{cyc}}}}} {{\tau _{{\rm{rel,}}}}_j}}} \right.
 \kern-\nulldelimiterspace} {{\tau _{{\rm{rel,}}}}_j}}}} - 1} \right)}}} \right\}} 
\end{array}
\end{align}
where $\min \left( {{p_1},{p_2},...,{p_n}} \right) \le {p_{\rm{a}}} \le \max \left( {{p_1},{p_2},...,{p_n}} \right)$.

\vfill

\end{document}